\documentclass[a4paper,11pt]{article}
\pdfoutput=1
\usepackage{jheppub}
\usepackage{graphicx}
\usepackage{subfigure}
\usepackage{dcolumn}
\usepackage{verbatim}
\usepackage{amsmath}
\usepackage{amssymb}
\usepackage{float}
\usepackage[makeroom]{cancel}
\usepackage{multirow}
\usepackage{mathtools}
\usepackage[T1]{fontenc}
\usepackage{tikz}
\usetikzlibrary{calc,matrix}
\newcommand{\txm} {\text{T}\chi\text{M}}
\newcommand{\tpm} {\text{T}\phi\text{M}}
\newcommand{\om} {\omega}
\newcommand{\ob} {\bar{\omega}}
\newcommand{\tb} {\bar{3}}
\newcommand{\ab} {\bar{3^p}}
\newcommand{\bb} {\bar{3^q}}
\newcommand{\cb} {\bar{3^{r}}}
\newcommand{\xb} {\bar{6}}

\title{\boldmath Fully Constrained Majorana Neutrino Mass Matrices using $\Sigma(72\times3)$}

\author[a]{R.~Krishnan}

\affiliation[a]{University of Warwick,\\Coventry, CV4 7AL, UK}

\emailAdd{k.rama@warwick.ac.uk}

\abstract{In 2002, two neutrino mixing ansatze having trimaximally mixed middle ($\nu_2$) columns, namely tri-chi-maximal mixing ($\txm$) and tri-phi-maximal mixing ($\tpm$), were proposed. It was recently shown that $\txm$ with $\chi=\pm \frac{\pi}{16}$ as well as $\tpm$ with $\phi = \pm \frac{\pi}{16}$ leads to the solution, $\sin^2 \theta_{13} = \frac{2}{3} \sin^2 \frac{\pi}{16}$, consistent with the latest measurements of the reactor mixing angle, $\theta_{13}$. To obtain $\txm_{(\chi=\pm \frac{\pi}{16})}$ and $\tpm_{(\phi=\pm \frac{\pi}{16})}$, we utilised the type~I see-saw framework with fully constrained Majorana neutrino mass matrices. These mass matrices also resulted in a relation among the neutrino masses, $m_1:m_2:m_3=\frac{\left(2+\sqrt{2}\right)}{1+\sqrt{2(2+\sqrt{2})}}:1:\frac{\left(2+\sqrt{2}\right)}{-1+\sqrt{2(2+\sqrt{2})}}$. In this paper we construct a flavour model based on the discrete group $\Sigma(72\times3)$ and obtain the aforementioned results. A Majorana neutrino mass matrix (a symmetric $3\times3$ matrix with 6 complex degrees of freedom) is conveniently mapped into a flavon field transforming as the complex 6 dimensional representation of $\Sigma(72\times3)$. Specific vacuum alignments of the flavons are used to arrive at the desired mass matrices.}

\begin{document}


\maketitle
\flushbottom

\section{Introduction}

The neutrino mixing information is encapsulated in the unitary PMNS mixing matrix which, in the standard PDG parameterisation~\cite{PDG}, is given by
\begin{equation}
U_\text{PMNS}=\left(\begin{matrix}
c_{12} c_{13} &s_{12} c_{13} &s_{13}e^{-i\delta} \\
-s_{12} c_{23}-c_{12} s_{23} s_{13} e^{i\delta} &c_{12}c_{23} -s_{12}s_{23} s_{13} e^{i\delta} &s_{23} c_{13} \\
s_{12} s_{23}-c_{12} c_{23} s_{13}e^{i\delta} &-c_{12}s_{23}-s_{12}c_{23}s_{13}e^{i\delta} &c_{23}c_{13}
\end{matrix}\right)\left(\begin{matrix}
1 &0 &0 \\
0 &e^{i\frac{\alpha_{21}}{2}} &0 \\
0 &0 &e^{i\frac{\alpha_{31}}{2}}
\end{matrix}\right)
\label{eq:pmns}
\end{equation}
where $s_{ij}=\sin \theta_{ij}, c_{ij}=\cos \theta_{ij}$. The three mixing angles $\theta_{12}$ (solar angle), $\theta_{23}$ (atmospheric angle) and $\theta_{13}$ (reactor angle) along with the $CP$-violating complex phases (the Dirac phase, $\delta$, and the two Majorana phases, $\alpha_{21}$ and $\alpha_{31}$) parameterise $U_{PMNS}$. In comparison to the small mixing angles observed in the quark sector, the neutrino mixing angles are found to be relatively large~\cite{NeutrinoGlobalFit}:
\begin{align}
\sin^2 \theta_{12} &= 0.313_{-0.012}^{+0.013}\,,\label{eq:anglevalues1}\\
\sin^2 \theta_{23} &= 0.444_{-0.031}^{+0.036}\,\, \text{and} \,\, 0.600_{-0.026}^{+0.019}\,,\label{eq:anglevalues2}\\
\sin^2 \theta_{13} &= 0.0244_{-0.0019}^{+0.0020}\,. \label{eq:anglevalues3}
\end{align}
The values of the complex phases are unknown at present. Besides measuring the mixing angles, the neutrino oscillation experiments also proved that neutrinos are massive particles. These experiments measure the mass-squared-differences of the neutrinos and currently their values are known at about $3\%$ precision~\cite{NeutrinoGlobalFit},
\begin{gather}
\Delta m_{21}^2=75.0_{-1.7}^{+1.9}~\text{meV}^2,\label{eq:massvalues1}\\
|\Delta m_{31}^2|=2429_{-54}^{+55}~\text{meV}^2.\label{eq:massvalues2}
\end{gather}

Several mixing ansatze with a trimaximally mixed second column for $U_\text{PMNS}$, i.e.~$|U_{e2}|=|U_{\mu2}|=|U_{\tau2}|=\frac{1}{\sqrt{3}}$, were proposed during the early 2000s~\cite{TM, TBM, TXPM, Xing, Demo}. Here we briefly revisit two of those, the tri-chi-maximal mixing ($\txm$) and the tri-phi-maximal mixing ($\tpm$),\footnote{$TM_i$ ($TM^i$) has been proposed~\cite{TM2a, TM2b} as a nomenclature to denote the mixing matrices that preserve various rows (columns) of the tribimaximal mixing~\cite{TBM}. Under this notation, both $\txm$ and $\tpm$ fall under the category of $TM_2$. To be more specific, $TM_2$ which breaks $CP$ maximally is $\txm$ and $TM_2$ which conserves $CP$ is $\tpm$.} which are relevant to our model. They can be conveniently parameterised~\cite{TXPM} as follows   
\begin{align}
U_{\txm}&=\left(\begin{matrix}\sqrt{\frac{2}{3}}\cos \chi & \frac{1}{\sqrt{3}} & \sqrt{\frac{2}{3}}\sin \chi\\
-\frac{\cos \chi}{\sqrt{6}}-i\frac{\sin \chi}{\sqrt{2}} & \frac{1}{\sqrt{3}} & i\frac{\cos \chi}{\sqrt{2}}-\frac{\sin \chi}{\sqrt{6}}\\
-\frac{\cos \chi}{\sqrt{6}}+i\frac{\sin \chi}{\sqrt{2}} & \frac{1}{\sqrt{3}} & -i\frac{\cos \chi}{\sqrt{2}}-\frac{\sin \chi}{\sqrt{6}}
\end{matrix}\right),\label{eq:txmform}\\
U_{\tpm}&=\left(\begin{matrix}\sqrt{\frac{2}{3}}\cos \phi & \frac{1}{\sqrt{3}} & \sqrt{\frac{2}{3}}\sin \phi\\
-\frac{\cos \phi}{\sqrt{6}}-\frac{\sin \phi}{\sqrt{2}} & \frac{1}{\sqrt{3}} & \frac{\cos \phi}{\sqrt{2}}-\frac{\sin \phi}{\sqrt{6}}\\
-\frac{\cos \phi}{\sqrt{6}}+\frac{\sin \phi}{\sqrt{2}} & \frac{1}{\sqrt{3}} & -\frac{\cos \phi}{\sqrt{2}}-\frac{\sin \phi}{\sqrt{6}}
\end{matrix}\right)\label{eq:tpmform}.
\end{align}
Both $\txm$ and $\tpm$ have one free parameter each ($\chi$ and $\phi$) which directly corresponds to the reactor mixing angle, $\theta_{13}$, through the $U_{e3}$ elements of the mixing matrices. The three mixing angles and the Dirac $CP$ phase obtained by relating Eq.~(\ref{eq:pmns}) with Eqs.~(\ref{eq:txmform},~\ref{eq:tpmform}) are shown in Table~\ref{tab:anglesandphase}.
{\renewcommand{\arraystretch}{1.6}
\begin{table}[H]
\begin{center}
\begin{tabular}{||c||c|c|c|c||}
\hline
\hline
	&$\sin^2 \theta_{13}$	&$\sin^2 \theta_{12}$	&$\sin^2 \theta_{23}$	&$\delta$	\\
\hline
\hline
$\txm$	&$\frac{2}{3} \sin^2 \chi$	&$\frac{1}{\left(3-2\sin^2 \chi\right)}$	&$\frac{1}{2}$	&$\pm\frac{\pi}{2}$	\\
\hline
$\tpm$	&$\frac{2}{3} \sin^2 \phi$	&$\frac{1}{\left(3-2\sin^2 \phi\right)}$	&$\frac{2 \sin^2 \left(\frac{2\pi}{3}+\phi\right)}{\left(3-2\sin^2 \phi\right)}$	&$0,~\pi$	\\
\hline
\hline
\end{tabular}
\end{center}
\caption{The standard PDG observables $\theta_{13}$, $\theta_{12}$, $\theta_{23}$ and $\delta$ in terms of the parameters $\chi$ and $\phi$. Note that the range of $\chi$ as well as $\phi$ is $-\frac{\pi}{2}$ to $+\frac{\pi}{2}$. In $\txm$ ($\tpm$), the parameter $\chi$ ($\phi$) being in the first and the fourth quadrant correspond to $\delta$ equal to $+\frac{\pi}{2}$ ($0$) and $-\frac{\pi}{2}$ ($\pi$) respectively.}
\label{tab:anglesandphase}
\end{table}}
\noindent In $\txm$, since $\delta = \pm \frac{\pi}{2}$, $CP$ violation is maximal for a given set of mixing angles. The Jarlskog $CP$ violating invariant~\cite{JCP1, JCP2, JCP3, JCP4, JCP5} in the context of $\txm$~\cite{TXPM} is given by 
\begin{equation}\label{eq:jcp}
J=\frac{\sin 2\chi}{6\sqrt{3}}.
\end{equation} 
On the other hand, $\tpm$ is $CP$ conserving, i.e.~$\delta = 0,~\pi$, and thus $J=0$. Since the reactor angle was discovered to be non-zero in 2012~\cite{DayaBay}, there has been a resurgence of interest~\cite{TM21, TM22, TM23, TM24, TM25, S4Paper, LIS, SteveFour, Delta16, Thomas} in $\txm$ and $\tpm$ and their equivalent forms\footnote{Any $CP$-conserving ($\delta = 0,~\pi$) mixing matrix with non-zero $\theta_{13}$ and trimaximally mixed $\nu_2$ column is equivalent to $\tpm$. Observationally they differ only with respect to the Majorana phases. Similarly any mixing matrix with $\delta = \pm \frac{\pi}{2}$, $\theta_{13}\neq0$ and trimaximal $\nu_2$ column is equivalent to $\txm$.}.  

Recently~\cite{LIS} it was shown that $\txm_{(\chi=\pm \frac{\pi}{16})}$ as well as $\tpm_{(\phi=\pm \frac{\pi}{16})}$ results in a reactor mixing angle, 
\begin{equation}
\begin{split}
\sin^2 \theta_{13} &= \frac{2}{3} \sin^2 \frac{\pi}{16} \\
&= 0.025,
\end{split}
\end{equation} 
consistent with the experimental data. The model was constructed in the Type-1 see-saw framework. Four cases of Majorana mass matrices were discussed:
\begin{align}
M_\text{Maj} & \propto \left(\begin{matrix}2-\sqrt{2} & 0 & \frac{1}{\sqrt{2}}\\
0 & 1 & 0\\
\frac{1}{\sqrt{2}} & 0 & 0
\end{matrix}\right), & M_\text{Maj} & \propto \left(\begin{matrix}0 & 0 & \frac{1}{\sqrt{2}}\\
0 & 1 & 0\\
\frac{1}{\sqrt{2}} & 0 & 2-\sqrt{2}
\end{matrix}\right),\label{eq:txmmat}\\
M_\text{Maj} & \propto \left(\begin{matrix}i+\frac{1-i}{\sqrt{2}} & 0 & 1-\frac{1}{\sqrt{2}}\\
0 & 1 & 0\\
1-\frac{1}{\sqrt{2}} & 0 & -i+\frac{1+i}{\sqrt{2}}
\end{matrix}\right), & M_\text{Maj} & \propto \left(\begin{matrix}-i+\frac{1+i}{\sqrt{2}} & 0 & 1-\frac{1}{\sqrt{2}}\\
0 & 1 & 0\\
1-\frac{1}{\sqrt{2}} & 0 & i+\frac{1-i}{\sqrt{2}}
\end{matrix}\right)\label{eq:tpmmat}
\end{align}
where $M_\text{Maj}$ is the coupling among the right-handed neutrino fields, i.e.~$\overline{(\nu_R)^c}M_\text{Maj} \nu_R$. In Ref.~\cite{LIS}, the mixing matrix was modelled in the form
\begin{equation}
U_\text{PMNS} = \mathcal{T} U_\nu
\end{equation}
where the $3\times3$ trimaximal contribution,
\begin{equation}
\mathcal{T}=\frac{1}{\sqrt{3}}\left(\begin{matrix}1 & 1 & 1\\
1 & \omega & \bar{\omega}\\
1 & \bar{\omega} & \omega
\end{matrix}\right) \quad \text{with} \quad \omega=e^{i\frac{2\pi}{3}}, \,\,\, \bar{\omega}=e^{\text{-}i\frac{2\pi}{3}},
\end{equation}
came from the charged-lepton sector. $U_\nu$, on the other hand, was the contribution from the neutrino sector. The four $U_\nu$s vis-a-vis the four Majorana neutrino mass matrices given in Eqs.~(\ref{eq:txmmat}) and Eqs.~(\ref{eq:tpmmat}), gave rise to $\txm_{(\chi=\pm \frac{\pi}{16})}$ and $\tpm_{(\phi=\pm \frac{\pi}{16})}$ respectively. All the four mass matrices, Eqs.~(\ref{eq:txmmat},~\ref{eq:tpmmat}), have the eigenvalues $\frac{1+\sqrt{2(2+\sqrt{2})}}{\left(2+\sqrt{2}\right)}$, $1$ and $\frac{-1+\sqrt{2(2+\sqrt{2})}}{\left(2+\sqrt{2}\right)}$. Due to the see-saw mechanism, the neutrino masses become inversely proportional to the eigenvalues of the Majorana mass matrices. As a result we obtained the mass relation
\begin{equation}\label{eq:numass}
m_1:m_2:m_3=\frac{\left(2+\sqrt{2}\right)}{1+\sqrt{2(2+\sqrt{2})}}:1:\frac{\left(2+\sqrt{2}\right)}{-1+\sqrt{2(2+\sqrt{2})}}\,\,.
\end{equation}
Using this mass relation and given the experimentally measured mass-squared differences, we also predicted the light neutrino mass to be around $25~\text{meV}$.

In this paper we use the discrete group $\Sigma(72\times3)$ to construct a flavon model that essentially reproduces the above results. Unlike the original paper~\cite{LIS} where the neutrino mass matrix was decomposed into a symmetric bi-product, here a single representation of the flavour group is used to build the symmetric mass matrix. A brief discussion of the group $\Sigma(72\times3)$ and its representations is provided in Section~2. Appendix~A contains further details such as the tensor product expansions of its various irreducible representations (irreps) and the corresponding Clebsch-Gordan (C-G) coefficients. In Section~3, we describe the model with its fermion and flavon field content in relation to these irreps. The flavons are assigned specific Vacuum Expectation Values (VEVs) to obtain the required mass matrices. How we may construct suitable flavon potentials to generate the given set of VEVs is demonstrated in Appendix~B. In Section~4, we obtain the phenomenological predictions and compare them with the current experimental data along with the possibility of further validation from future experiments. Finally the results are summarised in Section~5.

\section{The Group $\Sigma(72\times3)$ and its Representations}

Discrete groups have been used extensively in the description of flavour symmetries. Historically, the study of discrete groups can be traced back to the study of symmetries of geometrical objects. Tetrahedran, cube, octahedran, dodecahedran and icosahedran, which are the famous Platonic solids, were known to the ancient Greeks. These objects are the only regular polyhedra with congruent regular polygonal faces. Interestingly, the symmetry groups of the platonic solids are the most studied in the context of flavour symmetries too - $A_4$ (tetrahedron), $S_4$ (cube and its dual octahedron) and $A_5$ (dodecahedron and its dual icosahedron). These polyhedra live in the three-dimensional Euclidean space. In the context of flavour physics, it might be rewarding to study similar polyhedra that live in three-dimensional complex Hilbert space. In fact, five such complex polyhedra that correspond to the five Platonic solids exist as shown by Coxeter~\cite{Coxpoly}. They are $3\{3\}3\{3\}3$, $2\{3\}2\{4\}p$, $p\{4\}2\{3\}2$, $2\{4\}3\{3\}3$, $3\{3\}3\{4\}2$ where we have used the generalised schlafli symbols~\cite{Coxpoly} to represent the polyhedra. The polyhedron $3\{3\}3\{3\}3$ known as the Hessian polydehron can be thought of as the tetrahedron in the complex space. Its full symmetry group has 648 elements and is called $\Sigma(216\times3)$. Like the other discrete groups relevant in flavour symmetry, $\Sigma(216\times3)$ is also a subgroup of the continuous group $SU(3)$. 

The principal series of $\Sigma(216\times3)$~\cite{Sigma1} is given by
\begin{equation}
\{e\} \triangleleft Z_3 \triangleleft \Delta(27) \triangleleft \Delta(54) \triangleleft \Sigma(72\times3) \triangleleft \Sigma(216\times3).
\end{equation}
Our flavour symmetry group, $\Sigma(72\times3)$, is the maximal normal subgroup of $\Sigma(216\times3)$. So we get $\Sigma(216\times3)/\Sigma(72\times3)=Z_3$. Various details about the properties of the group $\Sigma(72\times3)$ and its representations can be found in Refs.~\cite{Sigma1, Sigma2, Smallgroup, SigmaHagedorn, Merle}. Note that $\Sigma(72\times3)$ is quite distinct from $\Sigma(216)$ which is defined using the relation $\Sigma(216\times3)/Z_3=\Sigma(216)$. In other words, $\Sigma(216\times3)$ forms the triple cover of $\Sigma(216)$. $\Sigma(216\times3)$ as well as $\Sigma(216)$ is sometimes referred to as the Hessian group. In terms of the GAP~\cite{GAP4} nomenclature, we have $\Sigma(216\times3)\equiv \text{SmallGroup(648,532)},\,$ $\Sigma(72\times3)\equiv \text{SmallGroup(216,88)}\,$ and $\,\Sigma(216)\equiv \text{SmallGroup(216,153)}$.

We find that, in the context of flavour physics and model building, $\Sigma(72\times3)$ has an appealing feature: it is the smallest group containing a complex three-dimensional representation whose tensor product with itself results in a complex six-dimensional representation, i.e.
\begin{equation}\label{eq:tensor0}
\boldsymbol{3}\otimes\boldsymbol{3}=\boldsymbol{6}\oplus\boldsymbol{\tb}.
\end{equation}
With a suitably chosen basis for $\boldsymbol{6}$ we get
\begin{equation}\label{eq:tensor1exp}
\boldsymbol{6}\equiv\left(\begin{matrix}\frac{1}{\sqrt{3}}\left(a_1 b_1 + a_2 b_2 + a_3 b_3\right)\\
	\frac{1}{\sqrt{6}} a_1 b_1 - \sqrt{\frac{2}{3}} a_2 b_2 + \frac{1}{\sqrt{6}} a_3 b_3\\
	\frac{1}{\sqrt{2}}\left(a_1 b_1 - a_3 b_3\right)\\
	\frac{1}{\sqrt{2}}\left(a_2 b_3 + a_3 b_2\right)\\
	\frac{1}{\sqrt{2}}\left(a_3 b_1 + a_1 b_3\right)\\
\frac{1}{\sqrt{2}}\left(a_1 b_2 + a_2 b_1\right)
\end{matrix}\right), \quad \quad \boldsymbol{\bar{3}} \equiv
\left(\begin{matrix}\frac{1}{\sqrt{2}}\left(a_2 b_3 - a_3 b_2\right)\\
	\frac{1}{\sqrt{2}}\left(a_3 b_1 - a_1 b_3\right)\\
\frac{1}{\sqrt{2}}\left(a_1 b_2 - a_2 b_1\right)
\end{matrix}\right)
\end{equation}
where $(a_1, a_2, a_3)^T$ and $(b_1, b_2, b_3)^T$ represent the first triplet and the second triplet respectively appearing in the LHS of Eq.~(\ref{eq:tensor0}). All the symmetric components of the tensor product together form the representation $\boldsymbol{6}$ and the antisymmetric components form $\boldsymbol{\tb}$. For the $SU(3)$ group it is well known that the tensor product of two $\boldsymbol{3}$s gives rise to a symmetric $\boldsymbol{6}$ and an antisymmetric $\boldsymbol{\tb}$. $\Sigma(72\times3)$ being a subgroup of $SU(3)$, of course, has its $\boldsymbol{6}$ and $\boldsymbol{\tb}$ embedded in the $\boldsymbol{6}$ and $\boldsymbol{\tb}$ of $SU(3)$.  

Consider the complex conjugation of Eq.~(\ref{eq:tensor0}), i.e.~$\boldsymbol{\tb}\otimes\boldsymbol{\tb}=\boldsymbol{\xb}\oplus\boldsymbol{3}$. Let the right-handed neutrinos form a triplet, $\nu_R=(\nu_{R1},\nu_{R2},\nu_{R3})^T$, which transforms as a $\boldsymbol{\tb}$. A symmetric (and also Lorentz invariant) combination of two such triplets leads to a sextet, $X_\nu$, which transforms as a $\boldsymbol{\xb}$,
\begin{equation}\label{eq:Xnu}
X_\nu = \left(\begin{matrix}\frac{1}{\sqrt{3}}\left(\nu_{R1}.\nu_{R1} + \nu_{R2}.\nu_{R2} + \nu_{R3}.\nu_{R3}\right)\\
	\frac{1}{\sqrt{6}} \nu_{R1}.\nu_{R1} - \sqrt{\frac{2}{3}} \nu_{R2}.\nu_{R2} + \frac{1}{\sqrt{6}} \nu_{R3}.\nu_{R3}\\
	\frac{1}{\sqrt{2}}\left(\nu_{R1}.\nu_{R1} - \nu_{R3}.\nu_{R3}\right)\\
	\frac{1}{\sqrt{2}}\left(\nu_{R2}.\nu_{R3} + \nu_{R3}.\nu_{R2}\right)\\
	\frac{1}{\sqrt{2}}\left(\nu_{R3}.\nu_{R1} + \nu_{R1}.\nu_{R3}\right)\\
\frac{1}{\sqrt{2}}\left(\nu_{R1}.\nu_{R2} + \nu_{R2}.\nu_{R1}\right)
\end{matrix}\right)\equiv \boldsymbol{\xb}
\end{equation}
where $\nu_i.\nu_j$ is the Lorentz invariant product of the right-handed neutrino Weyl spinors. We may couple $X_\nu$ to a flavon field $\phi=(\phi_1,\phi_2,\phi_3,\phi_4,\phi_5,\phi_6)^T$ which transforms as a $\boldsymbol{6}$ to construct the invariant term
\begin{equation}\label{eq:Tnu}
X_\nu^T \phi = \left(\begin{matrix}\nu_{R1}\\
	\nu_{R2}\\
\nu_{R3}
\end{matrix}\right)^T \left(\begin{matrix}\frac{\phi_1}{\sqrt{3}}+\frac{\phi_2}{\sqrt{6}}+\frac{\phi_3}{\sqrt{2}} & \frac{\phi_6}{\sqrt{2}} & \frac{\phi_5}{\sqrt{2}}\\
       \frac{\phi_6}{\sqrt{2}} & \frac{\phi_1}{\sqrt{3}}-\frac{\sqrt{2}\phi_2}{\sqrt{3}} & \frac{\phi_4}{\sqrt{2}}\\
       \frac{\phi_5}{\sqrt{2}} & \frac{\phi_4}{\sqrt{2}} & \frac{\phi_1}{\sqrt{3}}+\frac{\phi_2}{\sqrt{6}}-\frac{\phi_3}{\sqrt{2}}
\end{matrix}\right)\left(\begin{matrix}\nu_{R1}\\
	\nu_{R2}\\
\nu_{R3}
\end{matrix}\right).
\end{equation}
In general, the $3\times3$ Majorana mass matrix is symmetric and has six complex degrees of freedom. Therefore, using Eq.~(\ref{eq:Tnu}), any required mass matrix can be obtained through a suitably chosen Vacuum Expectation Value (VEV) for the flavon field. Constructing the symmetric Majorana neutrino mass matrix with the help of flavon sextets has been attempted before, eg. scalar fields transforming as the antisextets of $SU(3)_L$ are used in Refs.~\cite{Long1, Long2}.

To describe the representation theory of $\Sigma(72\times3)$ we largely follow Ref.~\cite{Sigma1}. $\Sigma(72\times3)$ can be constructed using four generators, namely $C$, $E$, $V$ and $X$~\cite{Sigma1}. For the three-dimensional representation, we have
\begin{equation}\label{eq:gen3}
C \equiv
\left(\begin{matrix}1 & 0 & 0\\
       0 & \om & 0\\
       0 & 0 & \ob
\end{matrix}\right), \quad E \equiv
\left(\begin{matrix}0 & 1 & 0\\
       0 & 0 & 1\\
       1 & 0 & 0
\end{matrix}\right), \quad V\equiv
-\frac{i}{\sqrt{3}}\left(\begin{matrix}1 & 1 & 1\\
       1 & \om & \ob\\
       1 & \ob & \om
\end{matrix}\right), \quad X\equiv
-\frac{i}{\sqrt{3}}\left(\begin{matrix}1 & 1 & \ob\\
       1 & \om & \om\\       
       \om & 1 & \om
       \end{matrix}\right).
\end{equation}
The characters of the representations of $\Sigma(72\times3)$ are given in Table~\ref{tab:charactertable}. From the character table it is easy to infer that the one-dimensional representations $\boldsymbol{1^p}$, $\boldsymbol{1^q}$ and $\boldsymbol{1^r}$ involve a multiplication with $\pm1$ only. For these representations, the generators $C$, $E$, $V$ and $X$ are given by
\begin{align}
\boldsymbol{1^p}:\quad \quad &C \equiv 1, \quad E \equiv 1, \quad V \equiv -1, \quad X \equiv 1 \label{eq:gen1p},\\ 
\boldsymbol{1^q}:\quad \quad &C \equiv 1, \quad E \equiv 1, \quad V \equiv 1, \quad X \equiv -1 \label{eq:gen1q},\\ 
\boldsymbol{1^r}:\quad \quad &C \equiv 1, \quad E \equiv 1, \quad V \equiv -1, \quad X \equiv -1 \label{eq:gen1r}.
\end{align}
The representations $\boldsymbol{1^p}$, $\boldsymbol{1^q}$ and $\boldsymbol{1^r}$ along with the representation $\boldsymbol{3}$ can be used to construct $\boldsymbol{3^p}$, $\boldsymbol{3^q}$ and $\boldsymbol{3^r}$:
\begin{equation}\label{eq:3times1}
\boldsymbol{3^p} = \boldsymbol{1^p} \otimes \boldsymbol{3}, \quad \boldsymbol{3^q} = \boldsymbol{1^q} \otimes \boldsymbol{3}, \quad \boldsymbol{3^r} = \boldsymbol{1^r} \otimes \boldsymbol{3}.
\end{equation}
For $\boldsymbol{3^p}$, $\boldsymbol{3^q}$ and $\boldsymbol{3^r}$, we use the basis defined using the generator matrices given in Eqs.~(\ref{eq:gen3}) multiplied with $\pm1$ in accordance with Eqs.~(\ref{eq:gen1p},~\ref{eq:gen1q},~\ref{eq:gen1r},~\ref{eq:3times1}). Tensor product expansions of various representations relevant to our model along with the $SU(3)$ embeddings (branching rules) are given in the Appendix~A. We have also provided the C-G coefficients and the generator matrices in the bases corresponding to those coefficients.

\begin{table}[H]
\begin{center}
\scriptsize
\begin{tabular}{||c||c c c|c|c c c|c c c|c c c|c c c||}
\hline
\hline
$\Sigma(72\times3)$&$C_1$&$C_2$&$C_3$&$C_4$&$C_5$&$C_6$&$C_7$&$C_8$&$C_9$&$C_{10}$&$C_{11}$&$C_{12}$&$C_{13}$&$C_{14}$&$C_{15}$&$C_{16}$\\
$\#C_k$&$1$&$1$&$1$&$24$&$9$&$9$&$9$&$18$&$18$&$18$&$18$&$18$&$18$&$18$&$18$&$18$\\
$ord(C_k)$&$1$&$3$&$3$&$3$&$2$&$6$&$6$&$4$&$12$&$12$&$4$&$12$&$12$&$4$&$12$&$12$\\
\hline
\hline
$\boldsymbol{1}$&$1$&$1$&$1$		&$1$&$1$&$1$&$1$		&$1$&$1$&$1$		&$1$&$1$&$1$		&$1$&$1$&$1$\\

\hline
$\boldsymbol{1^p}$&$1$&$1$&$1$		&$1$&$1$&$1$&$1$		&$-1$&$-1$&$-1$		&$1$&$1$&$1$		&$-1$&$-1$&$-1$\\
$\boldsymbol{1^q}$&$1$&$1$&$1$		&$1$&$1$&$1$&$1$		&$1$&$1$&$1$		&$-1$&$-1$&$-1$		&$-1$&$-1$&$-1$\\
$\boldsymbol{1^r}$&$1$&$1$&$1$		&$1$&$1$&$1$&$1$		&$-1$&$-1$&$-1$		&$-1$&$-1$&$-1$		&$1$&$1$&$1$\\
\hline
$\boldsymbol{2}$&$2$&$2$&$2$		&$2$&$-2$&$-2$&$-2$		&$0$&$0$&$0$		&$0$&$0$&$0$		&$0$&$0$&$0$\\
\hline
$\boldsymbol{3}$&$3$&$3\om$&$3\ob$	&$0$&$-1$&$-\om$&$-\ob$		&$1$&$\om$&$\ob$	&$1$&$\om$&$\ob$	&$1$&$\om$&$\ob$\\
\hline
$\boldsymbol{3^p}$&$3$&$3\om$&$3\ob$	&$0$&$-1$&$-\om$&$-\ob$		&$-1$&$-\om$&$-\ob$	&$1$&$\om$&$\ob$	&$-1$&$-\om$&$-\ob$\\
$\boldsymbol{3^q}$&$3$&$3\om$&$3\ob$	&$0$&$-1$&$-\om$&$-\ob$		&$1$&$\om$&$\ob$	&$-1$&$-\om$&$-\ob$	&$-1$&$-\om$&$-\ob$\\
$\boldsymbol{3^r}$&$3$&$3\om$&$3\ob$	&$0$&$-1$&$-\om$&$-\ob$		&$-1$&$-\om$&$-\ob$	&$-1$&$-\om$&$-\ob$	&$1$&$\om$&$\ob$\\
\hline
$\boldsymbol{\tb}$&$3$&$3\ob$&$3\om$	&$0$&$-1$&$-\ob$&$-\om$		&$1$&$\ob$&$\om$	&$1$&$\ob$&$\om$	&$1$&$\ob$&$\om$\\
\hline
$\boldsymbol{\ab}$&$3$&$3\ob$&$3\om$&$0$&$-1$&$-\ob$&$-\om$	&$-1$&$-\ob$&$-\om$	&$1$&$\ob$&$\om$	&$-1$&$-\ob$&$-\om$\\
$\boldsymbol{\bb}$&$3$&$3\ob$&$3\om$&$0$&$-1$&$-\ob$&$-\om$	&$1$&$\ob$&$\om$	&$-1$&$-\ob$&$-\om$	&$-1$&$-\ob$&$-\om$\\
$\boldsymbol{\cb}$&$3$&$3\ob$&$3\om$&$0$&$-1$&$-\ob$&$-\om$	&$-1$&$-\ob$&$-\om$	&$-1$&$-\ob$&$-\om$	&$1$&$\ob$&$\om$\\
\hline
$\boldsymbol{6}$&$6$&$6\ob$&$6\om$&$0$&$2$&$2\ob$&$2\om$		&$0$&$0$&$0$		&$0$&$0$&$0$		&$0$&$0$&$0$\\
$\boldsymbol{\xb}$&$6$&$6\om$&$6\ob$	&$0$&$2$&$2\om$&$2\ob$		&$0$&$0$&$0$		&$0$&$0$&$0$		&$0$&$0$&$0$\\
\hline
$\boldsymbol{8}$&$8$&$8$&$8$		&$-1$&$0$&$0$&$0$		&$0$&$0$&$0$		&$0$&$0$&$0$		&$0$&$0$&$0$\\
\hline
\hline
\end{tabular}
\end{center}
\caption{Character table of $\Sigma(72\times3)$.}
\label{tab:charactertable}
\end{table} 

\section{The Model}
\addtocontents{toc}{\protect\setcounter{tocdepth}{1}}

In this paper we construct our model in the Standard Model framework with the addition of heavy right-handed neutrinos. Through the type~I see-saw mechanism, light Majorana neutrinos are produced. The fermion and flavon content of the model with the representations to which they belong is given in Table~\ref{tab:flavourcontent}. The Standard Model Higgs field is assigned to the trivial (singlet) representation of $\Sigma(72\times3)$.

\begin{table}[H]
\begin{center}
\begin{tabular}{|c|c c c c c c c c c|}
\hline
	&$e_R$	&$\mu_R$&$\tau_R$&$L$	&$\nu_R$&$\phi_e$&$\phi_\mu$&$\phi_\tau$&$\phi$\\
\hline

$\Sigma(72\times3)$	&$\boldsymbol{1^p}$&$\boldsymbol{1^q}$&$\boldsymbol{1^r}$&$\boldsymbol{\tb}$&$\boldsymbol{\tb}$&$\boldsymbol{\ab}$&$\boldsymbol{\bb}$&$\boldsymbol{\cb}$&$\boldsymbol{6}$\\
\hline
\end{tabular}
\end{center}
\caption{The flavour structure of the model. The three families of the left-handed-weak-isospin lepton doublets form the triplet $L$ and the three right-handed heavy neutrinos form the triplet $\nu_R$. The flavons $\phi_e$, $\phi_\mu$, $\phi_\tau$ and $\phi$, are scalar fields and are gauge invariants. On the other hand, they transform non-trivially under the flavour group.}
\label{tab:flavourcontent}
\end{table}

For the charged leptons, we obtain the mass term
\begin{equation}\label{eq:mclept}
\left(y_e L^\dagger e_R \frac{\phi_e}{\Lambda}+y_\mu L^\dagger \mu_R \frac{\phi_\mu}{\Lambda}+y_\tau L^\dagger \tau_R \frac{\phi_\tau}{\Lambda}\right)H+H.C.
\end{equation}
where $H$ is the Standard Model Higgs, $\Lambda$ is the cut-off scale and $y_i$ are the coupling constants. The VEV of the Higgs, $(0, h_o)$, breaks the weak gauge symmetry. For the flavons $\phi_e$, $\phi_\mu$ and $\phi_\tau$, we assign the vacuum alignments\footnote{Refer to Appendix~B for the details of the flavon potential that leads to these VEVs.} 
\begin{equation}\label{eq:leptvev}
\langle\phi_e\rangle= \frac{i}{\sqrt{3}}(1,1,1),\quad\langle\phi_\mu\rangle= \frac{i}{\sqrt{3}}(1,\ob,\om),\quad \langle\phi_\tau\rangle= \frac{i}{\sqrt{3}}(1,\om,\ob).
\end{equation}
As a result of these vacuum alignments we get the following charged-lepton mass term
\begin{equation}\label{eq:leptcontrib}
\left(\begin{matrix} e_L\\
\mu_L\\
\tau_L
\end{matrix}\right)^\dagger V^\dagger \left(\begin{matrix} m_e & 0 & 0\\
0 & m_\mu & 0\\
0 & 0 & m_\tau
\end{matrix}\right)\left(\begin{matrix} e_R\\
\mu_R\\
\tau_R
\end{matrix}\right)
\end{equation}
where $m_e=\frac{y_e h_o}{\Lambda}$ etc. $V$ is the $3\times3$ trimaximal matrix and is one of the generators of $\Sigma(72\times3)$ as given in Eqs.~(\ref{eq:gen3}).

Now, we write the Dirac mass term for the neutrinos:
\begin{equation}\label{eq:mdl2dirac}
2 y_w L^\dagger \nu_R \tilde{H}+H.C.
\end{equation}
where $\tilde{H}$ is the conjugate Higgs and $y_w$ is the coupling constant. With the help of Eq.~(\ref{eq:Tnu}), we also write the Majorana mass term for the neutrinos:
\begin{equation}\label{eq:mnurnur}
y_G X_\nu^T\frac{\phi}{\Lambda}
\end{equation}
where $y_G$ is the coupling which gives rise to the heavy right-handed Majorana masses. Let $\langle\phi\rangle$ be the VEV acquired by the sextet flavon $\phi$, and let $\boldsymbol{ \langle\phi\rangle}$ be the corresponding $3\times 3$ symmetric matrix of the form given in Eq.~(\ref{eq:Tnu}). Combining the mass terms, Eq.~(\ref{eq:mdl2dirac}) and Eq.~(\ref{eq:mnurnur}), and using the VEVs of the Higgs and the flavon, we obtain the Dirac-Majorana mass matrix:
\begin{equation}
M=\left(\begin{matrix}0 & y_w h_o I\\
       y_w h_o I & \,\,y_G \frac{1}{\Lambda}\boldsymbol{ \langle\phi\rangle}
\end{matrix}\right).
\end{equation}
The $6\times6$ mass matrix $M$, forms the coupling
\begin{equation}
M_{ij} \,\nu_i.\nu_j \quad \text{with} \quad  \nu=\left(\begin{matrix}\nu_{L}^*\\
	\nu_{R}
\end{matrix}\right)
\end{equation}
where $\nu_L=(\nu_e,\nu_\mu,\nu_\tau)^T$ are the left-handed neutrino flavour eigenstates.

Since $y_w h_o$ is at the weak scale and $y_G$ is at the GUT scale, small neutrino masses are generated through the see-saw mechanism. It can be shown~\cite{Majorana} that the resulting effective see-saw mass matrix is of the form
\begin{equation}\label{eq:seesaw}
M_\text{ss}=-\left(y_w h_o\right)^2\left(y_G \frac{{\boldsymbol{ \langle\phi\rangle}}}{\Lambda}\right)^{-1}.
\end{equation}
From Eq.~(\ref{eq:seesaw}) it is clear that the see-saw mechanism makes the light neutrino masses inversely proportional to the eigenvalues of the matrix $\boldsymbol{ \langle\phi\rangle}$. As foreseen in the Introduction, we now construct the four cases of the mass matrices, Eqs.~(\ref{eq:txmmat},~\ref{eq:tpmmat}), all of which resulting in the neutrino mass relation, Eq.~(\ref{eq:numass}). To achieve this we choose suitable vacuum alignments\footnote{Refer to Appendix~B for the details of the flavon potentials that lead to these VEVs.} for the sextet flavon $\phi$. 

\subsection{$\txm_{(\chi=+\frac{\pi}{16})}$} 

Here we assign the vacuum alignment
\begin{equation}\label{eq:vevtxmp}
\langle\phi\rangle = \left(\frac{3-\sqrt{2}}{\sqrt{3}},-\frac{1}{\sqrt{3}},-1+\sqrt{2},0,1,0\right).
\end{equation}
Using the symmetric matrix form of the sextet given in Eq.~(\ref{eq:Tnu}), we obtain
\begin{equation}\label{eq:vevsym1}
\boldsymbol{ \langle\phi\rangle} =\left(\begin{matrix}2-\sqrt{2} & 0 & \frac{1}{\sqrt{2}}\\
0 & 1 & 0\\
\frac{1}{\sqrt{2}} & 0 & 0
\end{matrix}\right).
\end{equation}
Diagonalising the corresponing effective see-saw mass matrix $M_{ss}$, Eq.~(\ref{eq:seesaw}), we get
\begin{equation}\label{eq:nucontrib}
U_\nu^\dagger M_{ss} U_\nu^* = \left(y_w h_o\right)^2\frac{\Lambda}{y_G} \text{Diag}\left({\textstyle\frac{\left(2+\sqrt{2}\right)}{1+\sqrt{2(2+\sqrt{2})}},1,\frac{\left(2+\sqrt{2}\right)}{-1+\sqrt{2(2+\sqrt{2})}}}\right)
\end{equation}
leading to the neutrino mass relation, Eq.~(\ref{eq:numass}). The unitary matrix $U_\nu$ is given by
\begin{equation}\label{eq:unu1}
U_\nu = i \left(\begin{matrix}\cos\left(\frac{3\pi}{16}\right) & 0 & -i\sin\left(\frac{3\pi}{16}\right)\\
0 & 1 & 0\\
\sin\left(\frac{3\pi}{16}\right) & 0 & i\cos\left(\frac{3\pi}{16}\right)
\end{matrix}\right).
\end{equation}
The product of the contribution from the charged-lepton sector i.e.~$V$ from Eqs.~(\ref{eq:leptcontrib},~\ref{eq:gen3}) and the contribution from the neutrino sector i.e.~$U_\nu$ from Eqs.~(\ref{eq:nucontrib},~\ref{eq:unu1}) results in the $\txm_{(\chi=+\frac{\pi}{16})}$ mixing:
\begin{equation}\label{eq:mix1}
U_\text{PMNS}=V.U_\nu=\left(\begin{matrix}1 & 0 & 0\\
0 & \om & 0\\
0 & 0 & \ob
\end{matrix}\right).\left(\begin{matrix}\sqrt{\frac{2}{3}}\cos \chi & \frac{1}{\sqrt{3}} & \sqrt{\frac{2}{3}}\sin \chi\\
-\frac{\cos \chi}{\sqrt{6}}-i\frac{\sin \chi}{\sqrt{2}} & \frac{1}{\sqrt{3}} & i\frac{\cos \chi}{\sqrt{2}}-\frac{\sin \chi}{\sqrt{6}}\\
-\frac{\cos \chi}{\sqrt{6}}+i\frac{\sin \chi}{\sqrt{2}} & \frac{1}{\sqrt{3}} & -i\frac{\cos \chi}{\sqrt{2}}-\frac{\sin \chi}{\sqrt{6}}
\end{matrix}\right).\left(\begin{matrix}1 & 0 & 0\\
0 & 1 & 0\\
0 & 0 & i
\end{matrix}\right)
\end{equation}
with $\chi=+\frac{\pi}{16}$.

\subsection{$\txm_{(\chi=-\frac{\pi}{16})}$} 
Here we assign the vacuum alignment
\begin{equation}\label{eq:vevtxmm}
\langle\phi\rangle = \left(\frac{3-\sqrt{2}}{\sqrt{3}},-\frac{1}{\sqrt{3}},1-\sqrt{2},0,1,0\right)
\end{equation}
resulting in the symmetric matrix
\begin{equation}\label{eq:vevsym2}
\boldsymbol{ \langle\phi\rangle}=\left(\begin{matrix}0 & 0 & \frac{1}{\sqrt{2}}\\
0 & 1 & 0\\
\frac{1}{\sqrt{2}} & 0 & 2-\sqrt{2}
\end{matrix}\right).
\end{equation}
In this case, the diagonalising matrix is
\begin{equation}
U_\nu =i \left(\begin{matrix}\cos\left(\frac{5\pi}{16}\right) & 0 & i\sin\left(\frac{5\pi}{16}\right)\\
0 & 1 & 0\\
\sin\left(\frac{5\pi}{16}\right) & 0 & -i\cos\left(\frac{5\pi}{16}\right)
\end{matrix}\right)
\end{equation}
and the corresponding mixing matrix is
\begin{equation}
U_\text{PMNS}=V.U_\nu=\left(\begin{matrix}1 & 0 & 0\\
0 & \om & 0\\
0 & 0 & \ob
\end{matrix}\right).\left(\begin{matrix}\sqrt{\frac{2}{3}}\cos \chi & \frac{1}{\sqrt{3}} & \sqrt{\frac{2}{3}}\sin \chi\\
-\frac{\cos \chi}{\sqrt{6}}-i\frac{\sin \chi}{\sqrt{2}} & \frac{1}{\sqrt{3}} & i\frac{\cos \chi}{\sqrt{2}}-\frac{\sin \chi}{\sqrt{6}}\\
-\frac{\cos \chi}{\sqrt{6}}+i\frac{\sin \chi}{\sqrt{2}} & \frac{1}{\sqrt{3}} & -i\frac{\cos \chi}{\sqrt{2}}-\frac{\sin \chi}{\sqrt{6}}
\end{matrix}\right).\left(\begin{matrix}1 & 0 & 0\\
0 & 1 & 0\\
0 & 0 & -i
\end{matrix}\right)\label{eq:mix2}
\end{equation}
with $\chi=-\frac{\pi}{16}$.

\subsection{$\tpm_{(\phi=+\frac{\pi}{16})}$} 
Here we assign the vacuum alignment
\begin{equation}\label{eq:vevtpmp}
\langle\phi\rangle = \left(\frac{1+\sqrt{2}}{\sqrt{3}},\frac{1-\sqrt{2}}{\sqrt{3}}, -i \left(1-\sqrt{2}\right),0, -1+\sqrt{2}, 0\right)
\end{equation}
resulting in the symmetric matrix
\begin{equation}\label{eq:vevsym3}
\boldsymbol{ \langle\phi\rangle}=\left(\begin{matrix}i+\frac{1-i}{\sqrt{2}} & 0 & 1-\frac{1}{\sqrt{2}}\\
0 & 1 & 0\\
1-\frac{1}{\sqrt{2}} & 0 & -i+\frac{1+i}{\sqrt{2}}
\end{matrix}\right).
\end{equation}
In this case, the diagonalising matrix is
\begin{equation}
U_\nu =i \left(\begin{matrix}\frac{1}{\sqrt{2}} e^{-i\frac{\pi}{16}} & 0 & -\frac{1}{\sqrt{2}} e^{-i\frac{\pi}{16}}\\
0 & 1 & 0\\
\frac{1}{\sqrt{2}} e^{i\frac{\pi}{16}} & 0 & \frac{1}{\sqrt{2}} e^{i\frac{\pi}{16}}
\end{matrix}\right)
\end{equation}
and the corresponding mixing matrix is
\begin{equation}\label{eq:mix3}
U_\text{PMNS}=V.U_\nu=\left(\begin{matrix}1 & 0 & 0\\
0 & \om & 0\\
0 & 0 & \ob
\end{matrix}\right).\left(\begin{matrix}\sqrt{\frac{2}{3}}\cos \phi & \frac{1}{\sqrt{3}} & \sqrt{\frac{2}{3}}\sin \phi\\
-\frac{\cos \phi}{\sqrt{6}}-\frac{\sin \phi}{\sqrt{2}} & \frac{1}{\sqrt{3}} & \frac{\cos \phi}{\sqrt{2}}-\frac{\sin \phi}{\sqrt{6}}\\
-\frac{\cos \phi}{\sqrt{6}}+\frac{\sin \phi}{\sqrt{2}} & \frac{1}{\sqrt{3}} & -\frac{\cos \phi}{\sqrt{2}}-\frac{\sin \phi}{\sqrt{6}}
\end{matrix}\right).\left(\begin{matrix}1 & 0 & 0\\
0 & 1 & 0\\
0 & 0 & i
\end{matrix}\right)
\end{equation}
with $\phi=+\frac{\pi}{16}$.

\subsection{$\tpm_{(\phi=-\frac{\pi}{16})}$} 
Here we assign the vacuum alignment
\begin{equation}\label{eq:vevtpmm}
\langle\phi\rangle = \left(\frac{1+\sqrt{2}}{\sqrt{3}},\frac{1-\sqrt{2}}{\sqrt{3}}, i (1-\sqrt{2}),0, -1+\sqrt{2}, 0\right)
\end{equation}
resulting in the symmetric matrix
\begin{equation}\label{eq:vevsym4}
\boldsymbol{ \langle\phi\rangle}=\left(\begin{matrix}-i+\frac{1+i}{\sqrt{2}} & 0 & 1-\frac{1}{\sqrt{2}}\\
0 & 1 & 0\\
1-\frac{1}{\sqrt{2}} & 0 & i+\frac{1-i}{\sqrt{2}}
\end{matrix}\right).
\end{equation}
In this case, the diagonalising matrix is
\begin{equation}
U_\nu =i \left(\begin{matrix}\frac{1}{\sqrt{2}} e^{i\frac{\pi}{16}} & 0 & \frac{1}{\sqrt{2}} e^{i\frac{\pi}{16}}\\
0 & 1 & 0\\
\frac{1}{\sqrt{2}} e^{-i\frac{\pi}{16}} & 0 & -\frac{1}{\sqrt{2}} e^{-i\frac{\pi}{16}}
\end{matrix}\right)
\end{equation}
and the corresponding mixing matrix is
\begin{equation}
U_\text{PMNS}=V.U_\nu=\left(\begin{matrix}1 & 0 & 0\\
0 & \om & 0\\
0 & 0 & \ob
\end{matrix}\right).\left(\begin{matrix}\sqrt{\frac{2}{3}}\cos \phi & \frac{1}{\sqrt{3}} & \sqrt{\frac{2}{3}}\sin \phi\\
-\frac{\cos \phi}{\sqrt{6}}-\frac{\sin \phi}{\sqrt{2}} & \frac{1}{\sqrt{3}} & \frac{\cos \phi}{\sqrt{2}}-\frac{\sin \phi}{\sqrt{6}}\\
-\frac{\cos \phi}{\sqrt{6}}+\frac{\sin \phi}{\sqrt{2}} & \frac{1}{\sqrt{3}} & -\frac{\cos \phi}{\sqrt{2}}-\frac{\sin \phi}{\sqrt{6}}
\end{matrix}\right).\left(\begin{matrix}1 & 0 & 0\\
0 & 1 & 0\\
0 & 0 & -i
\end{matrix}\right)\label{eq:mix4}
\end{equation}
with $\phi=-\frac{\pi}{16}$.

As stated earlier, the four cases, Eqs.~(\ref{eq:vevsym1},~\ref{eq:vevsym2},~\ref{eq:vevsym3},~\ref{eq:vevsym4}), result in the same neutrino mass relation Eq.~(\ref{eq:numass}).

\section{Predicted Observables}

For comparing our model with the neutrino oscillation experimental data, we use the global analysis done by the NuFIT group and their latest results reproduced in Eqs.~(\ref{eq:anglevalues1}-\ref{eq:massvalues2}). The prediction $\sin^2 \theta_{13} = \frac{2}{3} \sin^2 \frac{\pi}{16} = 0.025$,\footnote{Besides in Ref.~\cite{LIS}, this value was predicted in the context of $\Delta(6n^2)$ symmetry group in Ref.~\cite{Delta16} and later obtained in Ref.~\cite{Thomas}} is within $1\sigma$ errors. For the solar angle, using the formula given in Table~\ref{tab:anglesandphase}, we get 
\begin{equation}
\begin{split}
\sin^2 \theta_{12} &= \frac{1}{3-2\sin^2\left(\frac{\pi}{16}\right)}\\ 
&= 0.342\,.
\end{split}
\end{equation}
This has a small tension with the experimental value, but it is still within $3\sigma$ errors. For the atmospheric angle, $\txm$ predicts maximal mixing:
\begin{equation}
\sin^2 \theta_{23} = \frac{1}{2}\,.
\end{equation}
The NuFIT data as well as other global fits~\cite{Global1,Global1b,Global2} are showing a preference for non-maximal atmospheric mixing. As a result there has been a recent interest in the problem of octant degeneracy of $\theta_{23}$~\cite{Octant1,Octant2,Octant3}. $\tpm$ predicts this non-maximal scenario of atmospheric mixing. $\tpm_{(\phi=\frac{\pi}{16})}$ and $\tpm_{(\phi=-\frac{\pi}{16})}$ correspond to the first and the second octant solutions respectively. Using the formula for $\theta_{23}$ given in Table~{tab:anglesandphase}, we get 
\begin{gather}
\tpm_{(\phi=+\frac{\pi}{16})}:\,
\begin{split}
\quad \sin^2 \theta_{23} &= \frac{2\sin^2\left(\frac{2\pi}{3}+\frac{\pi}{16}\right)}{3-2\sin^2\left(\frac{\pi}{16}\right)} \\
&= 0.387\,,\\
\end{split}\\
\tpm_{(\phi=-\frac{\pi}{16})}:\,
\begin{split}
\quad \sin^2 \theta_{23} &= \frac{2\sin^2\left(\frac{2\pi}{3}-\frac{\pi}{16}\right)}{3-2\sin^2\left(\frac{\pi}{16}\right)}\\
&= 0.613\,.
\end{split}
\end{gather}
The dirac $CP$ phase, $\delta$, has not been measured yet. The discovery that the reactor mixing angle is not very small has raised the possibility of a relatively earlier measurement of $\delta$~\cite{MHCP1,MHCP2}. $\txm$ having $\delta=\pm\frac{\pi}{2}$ should lead to large observable $CP$ violating effects. Substituting $\chi=\pm\frac{\pi}{16}$ in Eq.~(\ref{eq:jcp}), our model gives
\begin{equation}
\begin{split}
J&=\pm\frac{\sin \frac{\pi}{8}}{6\sqrt{3}}\\
&= \pm0.0368.
\end{split}
\end{equation} 
which is about $40\%$ of the maximum value of the theoretical range, $-\frac{1}{6\sqrt{3}}\leq J\leq +\frac{1}{6\sqrt{3}}$.
On the other hand, $\tpm$, with $\delta=0,\,\pi$ and $J=0$, is $CP$ conserving. 

\begin{figure}[]
\begin{center}
\includegraphics[scale=1.0]{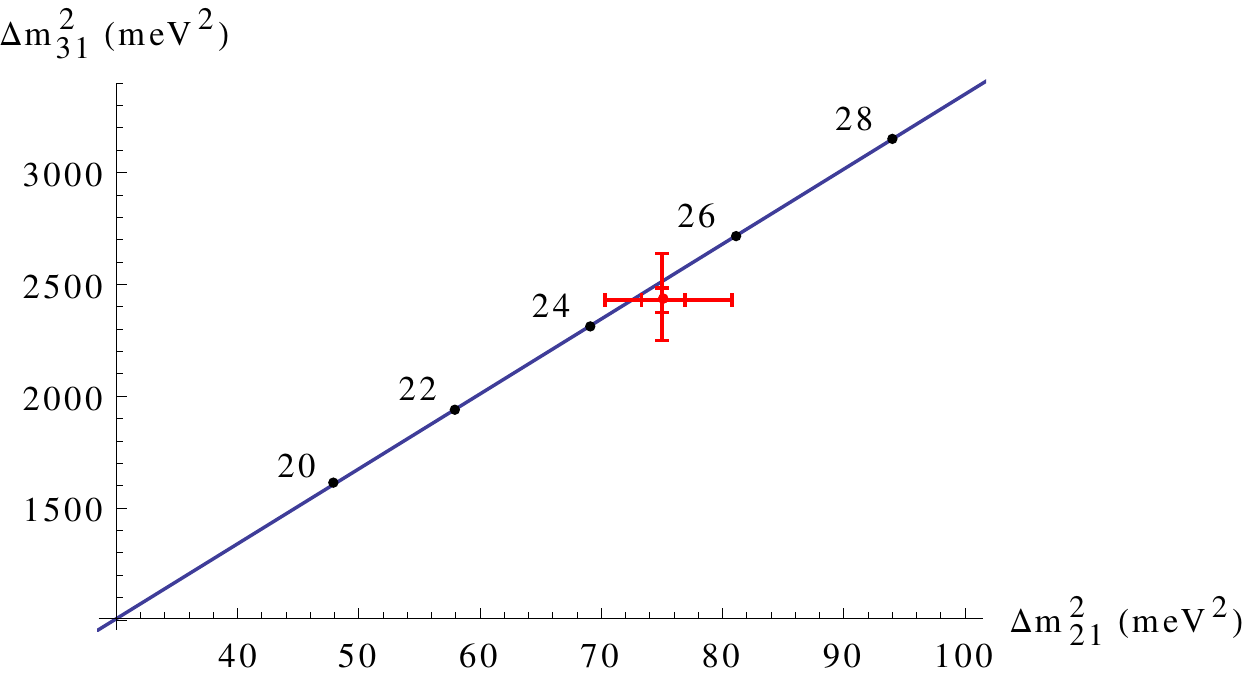}
\caption[$\Delta m_{31}^2$ vs $\Delta m_{21}^2$ plane]{$\Delta m_{31}^2$ vs $\Delta m_{21}^2$ plane. The straight line shows the neutrino mass relation Eq.~(\ref{eq:numass}). As a parametric plot, the line can be represented as $\Delta m_{21}^2=(r_{21}^2-1)m_1^2$ and $\Delta m_{31}^2=(r_{31}^2-1)m_1^2$ where $r_{21}=\frac{m_2}{m_1}=\frac{1+\sqrt{2(2+\sqrt{2})}}{\left(2+\sqrt{2}\right)}$ and $r_{31}=\frac{m_3}{m_1}=\frac{1+\sqrt{2(2+\sqrt{2})}}{-1+\sqrt{2(2+\sqrt{2})}}$ are the mass ratios obtained from Eq.~(\ref{eq:numass}). The parametric values of the light neutrino mass, $m_1$, (denoted by the black dots on the line) are in terms of meV. The red marking indicates the experimental best fit for $\Delta m_{21}^2$ and $\Delta m_{31}^2$ along with $1\sigma$ and $3\sigma$ errors.}
\label{fig:neutrinopredict2}
\end{center}
\end{figure}

From Figure~\ref{fig:neutrinopredict2}, it is clear that the neutrino mass relation~Eq.~(\ref{eq:numass}), is consistent with the measured mass-squared differences, Eqs.~(\ref{eq:massvalues1},~\ref{eq:massvalues2}). Here we have assumed the normal mass hierarchy. Using Eq.~(\ref{eq:numass}) and Eqs.~(\ref{eq:massvalues1},~\ref{eq:massvalues2}), we predict the neutrino masses:\footnote{The best fit values correspond to $\chi^2_\text{min}=1.1$ and the error ranges correspond to $\Delta\chi^2=1$ where $\chi^2=\displaystyle\sum_{{\displaystyle x}=\Delta m^2_{21}, \Delta m^2_{31}} \left(\frac{x_\text{model}-x_\text{expt}}{\sigma_{x\,\text{expt}}}\right)^2$ and $\Delta\chi^2=\chi^2-\chi^2_\text{min}$.}
\begin{gather}
m_1=24.77^{+0.20}_{-0.19}~\text{meV},\notag\\
m_2=26.22^{+0.21}_{-0.20}~\text{meV},\label{eq:m123}\\
m_3=55.49^{+0.45}_{-0.42}~\text{meV}.\notag
\end{gather}
Note that the mass relation Eq.~(\ref{eq:numass}), is incompatible with the inverted mass hierarchy. Considerable experimental studies are being conducted to determine the mass hierarchy~\cite{MHCP2,MH1,MH2,MH3} and we may expect a solution in the not-too-distant future. Observation of inverted hierarchy will obviously rule out the model.

Cosmological observations can give us limits on the sum of the neutrino masses. The strongest such limit has been set recently by the data collected using the Planck satellite~\cite{Planck}:
\begin{equation}
\sum_i m_i < 230~\text{meV}.
\end{equation}
Our predictions Eqs.~(\ref{eq:m123}), give a sum 
\begin{equation}
\sum_i m_i = 106.48^{+0.86}_{-0.81}~\text{meV}
\end{equation}
which is not far below the current cosmological limit. By including the data from future projects such as Polarbear and SKA, we may be able to lower the cosmological limit to around $100~\text{meV}$~\cite{FutureCosmo}. Such a result may support or rule out our model.

Neutrinoless double beta decay experiments seek to determine the Majorana nature of the neutrinos. These experiments have so far set limits on the effective electron neutrino mass~\cite{DBD} $|m_{\beta\beta}|$, where
\begin{align}
\begin{split}
m_{\beta\beta}&=m_1 U_{e1}^2+m_2 U_{e2}^2+m_3 U_{e3}^2\\
&=m_1 |U_{e1}|^2+m_2 |U_{e2}|^2 e^{i\alpha_{21}}+m_3 |U_{e3}|^2 e^{i\alpha_{31}}
\end{split}
\end{align}
with $U$ representing $U_\text{PMNS}$. In all the four mixing scenarios predicted by the model, Eqs.~(\ref{eq:mix1},~\ref{eq:mix2},~\ref{eq:mix3},~\ref{eq:mix4}), we have $|U_{e1}|=\sqrt{\frac{2}{3}}\cos \frac{\pi}{16}$, $|U_{e2}|=\frac{1}{\sqrt{3}}$ and $|U_{e3}|=\sqrt{\frac{2}{3}}\sin \frac{\pi}{16}$. Also, all of them result in the Majorana phases:\footnote{Note that both $+i$ and $-i$ appearing in the diagonal phase matrices in Eqs.~(\ref{eq:mix1},~\ref{eq:mix2},~\ref{eq:mix3},~\ref{eq:mix4}) correspond to $\alpha_{31}=\pi$}
\begin{equation}
\alpha_{21}=0, \quad \alpha_{31}=\pi. 
\end{equation}
Therefore the model predicts
\begin{equation}\label{eq:modelbb}
m_{\beta\beta}=\frac{2}{3}m_1\cos^2 \frac{\pi}{16}+\frac{1}{3}m_2-\frac{2}{3}m_3\sin^2 \frac{\pi}{16}.  
\end{equation}
Substituting the neutrino masses from Eqs.~(\ref{eq:m123}) in Eq.~(\ref{eq:modelbb}) we get 
\begin{equation}
m_{\beta\beta}=23.22^{+0.19}_{-0.18}~\text{meV}.  
\end{equation}

The most stringent upper bounds on the value of $|m_{\beta\beta}|$ have been set by Heidelberg-Moscow, Cuoricino, NEMO3 and GERDA experiments. These bounds are of the order of a few hundreds of meV. In future, experiments such as GERDA-2, CUORE and EXO can improve the bounds on $|m_{\beta\beta}|$ to a few tens~\cite{DBD} of meV and thus may support or rule out our model.

\section{Summary}

In this paper we utilise the group $\Sigma(72\times3)$ to construct fully-constrained Majorana mass matrices for the neutrinos. These mass matrices reproduce the results obtained in Ref.~\cite{LIS} i.e.~$\txm_{(\chi=\pm\frac{\pi}{16})}$ and $\tpm_{(\phi=\pm\frac{\pi}{16})}$ mixings along with the neutrino mass relation Eq.~(\ref{eq:numass}). The mixing observables as well as the neutrino mass relation are shown to be consistent with the experimental data. $\txm_{(\chi=\pm\frac{\pi}{16})}$ and $\tpm_{(\phi=\pm\frac{\pi}{16})}$ predict the Dirac $CP$ violating effect to be maximal (at fixed $\theta_{13}$) and null respectively. Using our neutrino mass relation in conjunction with the experimentally-observed neutrino mass-squared differences, we calculate the individual neutrino masses. We note that the neutrino mass relation Eq.~(\ref{eq:numass}) is incompatible with the inverted mass hierarchy. We also predict the effective electron neutrino mass for the neutrinoless double beta decay, $|m_{\beta\beta}|$. We briefly discuss the current status and future prospects of experimentally determining the neutrino observables leading to the confirmation or the falsification of our model. In the context of model building, we carry out an in-depth analysis of the representations of $\Sigma(72\times3)$ and develop the necessary groundwork to construct the flavon potentials satisfying the $\Sigma(72\times3)$ flavour symmetry. In the charged-lepton sector, we use three triplet flavons with a suitably chosen set of VEVs which provide a $3\times 3$ trimaximal contribution towards the PMNS mixing matrix. In the neutrino sector, we discuss four cases of Majorana mass matrices. The $\Sigma(72\times3)$ sextet acts as the most general placeholder for a fully constrained Majorana mass matrix. The intended mass matrices are obtained by assigning appropriate VEVs to the sextet flavon. It should be noted that we need additional symmetries to `explain' any specific texture in the mass matrix. Further research in this direction is a work in progress.

I would like to thank Paul Harrison and Bill Scott for helpful discussions. This work was supported by the UK Science and Technology Facilities Council (STFC). I acknowledge the hospitality of the Centre for Fundamental Physics (CfFP) at the Rutherford Appleton Laboratory and support from the University of Warwick. 

\section{Appendix A: Tensor Product Expansions of Irreps of $\Sigma(72\times3)$}

\begin{flalign}\label{eq:tensor288}
&\text{{\textit {\textbf {i}}})\,\,\,}\boldsymbol{3}\otimes\boldsymbol{3}=\boldsymbol{6}\oplus\boldsymbol{\tb}&
\end{flalign}
The generator matrices for the triplet representation were provided in Eq.~(\ref{eq:gen3}). We define the basis for the sextet representation using Eqs.~(\ref{eq:tensor1exp}). The resulting generator matrices are 
\begin{align}\label{eq:gen6}
\begin{split}
&C \equiv
\left(\begin{matrix}0 & -\frac{\ob}{\sqrt{2}} & \frac{i\ob}{\sqrt{2}} & 0 & 0 & 0\\
	-\frac{\ob}{\sqrt{2}} & \frac{\ob}{2} & \frac{i\ob}{2} & 0 & 0 & 0\\
	\frac{i\ob}{\sqrt{2}} & \frac{i\ob}{2} & -\frac{\ob}{2} & 0 & 0 & 0\\
	0 & 0 & 0 & 1 & 0 & 0\\
	0 & 0 & 0 & 0 & \ob & 0\\
	0 & 0 & 0 & 0 & 0 & \om
\end{matrix}\right), \quad E \equiv
\left(\begin{matrix}1 & 0 & 0 & 0 & 0 & 0\\
	0 & -\frac{1}{2} & \frac{\sqrt{3}}{2} & 0 & 0 & 0\\
	0 & -\frac{\sqrt{3}}{2} & -\frac{1}{2} & 0 & 0 & 0\\
	0 & 0 & 0 & 0 & 1 & 0\\
	0 & 0 & 0 & 0 & 0 & 1\\
	0 & 0 & 0 & 1 & 0 & 0
\end{matrix}\right),\\
&V\equiv
\left(\begin{matrix}-\frac{1}{3} & -\frac{1}{3\sqrt{2}} & -\frac{1}{\sqrt{6}} & -\sqrt{\frac{2}{3}} & 0 & 0\\
	-\frac{1}{3\sqrt{2}} & \frac{1}{3}+\frac{\om}{2} & -\frac{\om}{2\sqrt{3}} & 0 & \frac{\ob}{\sqrt{3}} & \frac{\om}{\sqrt{3}}\\
	-\frac{1}{\sqrt{6}} & -\frac{\om}{2\sqrt{3}} & -\frac{i\om}{2\sqrt{3}} & 0 & -\frac{i\ob}{\sqrt{3}} & \frac{i\om}{\sqrt{3}}\\
	-\sqrt{\frac{2}{3}} & 0 & 0 & \frac{1}{3} & \frac{1}{3} & \frac{1}{3}\\
	0 & \frac{\ob}{\sqrt{3}} & -\frac{i\ob}{\sqrt{3}} & \frac{1}{3} & \frac{\ob}{3} & \frac{\om}{3}\\
	0 & \frac{\om}{\sqrt{3}} & \frac{i\om}{\sqrt{3}} & \frac{1}{3} & \frac{\om}{3} & \frac{\ob}{3}
\end{matrix}\right),\\
&X\equiv
\left(\begin{matrix}\frac{\om}{3} & \frac{1}{3\sqrt{2}} & -\frac{1}{\sqrt{6}} & -\frac{i\om\sqrt{2}}{3} & -\frac{i\om\sqrt{2}}{3} & -\frac{i\sqrt{2}}{3}\\
	\frac{\ob}{3\sqrt{2}} & \frac{1}{2}+\frac{\om}{6} & -\frac{\ob}{2\sqrt{3}} & -\frac{i}{3} & \frac{i2}{3} & -\frac{i\ob}{3}\\
	\frac{\ob}{\sqrt{6}} & \frac{\ob}{2\sqrt{3}} & -\frac{i\ob}{2\sqrt{3}} & \frac{i}{\sqrt{3}} & 0 & -\frac{i\ob}{\sqrt{3}}\\
	\frac{i\ob\sqrt{2}}{3} & \frac{i}{3} & -\frac{i}{\sqrt{3}} & \frac{1}{3} & \frac{1}{3} & \frac{\om}{3}\\
	\frac{i\om\sqrt{2}}{3} & \frac{i\ob}{3} & \frac{i\ob}{\sqrt{3}} & \frac{1}{3} & \frac{\ob}{3} & \frac{\ob}{3}\\
	\frac{i\om\sqrt{2}}{3} & -\frac{i\ob2}{3} & 0 & \frac{\ob}{3} & \frac{1}{3} & \frac{\ob}{3}
\end{matrix}\right).
\end{split}
\end{align}

\newpage
\begin{flalign}\label{eq:tensor3pqr}
&\text{{\textit {\textbf {ii}}})\,\,\,}\boldsymbol{3^p}\otimes\boldsymbol{3^p}=\boldsymbol{6}\oplus\boldsymbol{\tb}, \quad \boldsymbol{3^q}\otimes\boldsymbol{3^q}=\boldsymbol{6}\oplus\boldsymbol{\tb}, \quad \boldsymbol{3^r}\otimes\boldsymbol{3^r}=\boldsymbol{6}\oplus\boldsymbol{\tb}&
\end{flalign}
We have already defined the bases of $\boldsymbol{3}$, $\boldsymbol{3^p}$, $\boldsymbol{3^q}$, $\boldsymbol{3^r}$ and $\boldsymbol{6}$. Since the representation matrices corresponding to $\boldsymbol{3}$, $\boldsymbol{3^p}$, $\boldsymbol{3^q}$ and $\boldsymbol{3^r}$ differ only with respect to the multiplication with $\pm1$, the C-G coefficients for the tensor product expansions in Eqs.~(\ref{eq:tensor3pqr}) are exactly those given in Eqs.~(\ref{eq:tensor1exp}). However that is not the case for tensor products involving different types of triplets, 
\begin{flalign}\label{eq:tensor3pqr2}
&\quad \,\,\,\boldsymbol{3^p}\otimes\boldsymbol{3^q}=\boldsymbol{6}\oplus\boldsymbol{\cb}, \quad \boldsymbol{3^q}\otimes\boldsymbol{3^r}=\boldsymbol{6}\oplus\boldsymbol{\ab}, \quad \boldsymbol{3^r}\otimes\boldsymbol{3^p}=\boldsymbol{6}\oplus\boldsymbol{\bb}.&
\end{flalign}
The C-G coefficients for the sextets appearing in the first, the second and the third tensor product expansions in Eqs.~(\ref{eq:tensor3pqr2}) are given by
\begin{equation}\label{eq:tensorpqrexp}
\boldsymbol{6}\equiv\left(\begin{matrix}\frac{1}{3}\left(a_1 b_1 + a_2 b_2 + a_3 b_3+\om\left(a_2 b_3+a_3 b_2+a_3 b_1 + a_1 b_3 +a_1 b_2 + a_2 b_1\right)\right)\\
	\frac{1}{3\sqrt{2}} a_1 b_1 - \frac{\sqrt{2}}{3} a_2 b_2 + \frac{1}{3\sqrt{2}} a_3 b_3+\frac{\om}{3\sqrt{2}} (a_2 b_3+a_3 b_2) - \frac{\om\sqrt{2}}{3} (a_3 b_1 + a_1 b_3) + \frac{\om}{3\sqrt{2}} (a_1 b_2 +a_2 b_1)\\
	\frac{1}{\sqrt{6}}\left(a_1 b_1 - a_3 b_3\right)+\frac{\om}{\sqrt{6}}\left(a_2 b_3 + a_3 b_2 - a_1 b_2 - a_2 b_1\right)\\
	\frac{\ob \sqrt{2}}{\sqrt{3}}a_1 b_1 - \frac{1}{\sqrt{6}}\left(a_2 b_3+a_3 b_2\right)\\
	\frac{\ob \sqrt{2}}{\sqrt{3}}a_2 b_2 - \frac{1}{\sqrt{6}}\left(a_3 b_1+a_1 b_3\right)\\
\frac{\ob \sqrt{2}}{\sqrt{3}}a_3 b_3 - \frac{1}{\sqrt{6}}\left(a_1 b_2+a_2 b_1\right),
\end{matrix}\right),
\end{equation}
\begin{equation}\label{eq:tensorqrexp}
\boldsymbol{6}\equiv\left(\begin{matrix}\frac{1}{3}\left(a_1 b_1 + a_2 b_2 + a_3 b_3+\ob\left(a_2 b_3+a_3 b_2+a_3 b_1 + a_1 b_3 +a_1 b_2 + a_2 b_1\right)\right)\\
	\frac{1}{3\sqrt{2}} a_1 b_1 - \frac{\sqrt{2}}{3} a_2 b_2 + \frac{1}{3\sqrt{2}} a_3 b_3+\frac{\ob}{3\sqrt{2}} (a_2 b_3+a_3 b_2) - \frac{\ob\sqrt{2}}{3} (a_3 b_1 + a_1 b_3) + \frac{\ob}{3\sqrt{2}} (a_1 b_2 +a_2 b_1)\\
	\frac{1}{\sqrt{6}}\left(a_1 b_1 - a_3 b_3\right)+\frac{\ob}{\sqrt{6}}\left(a_2 b_3 + a_3 b_2 - a_1 b_2 - a_2 b_1\right)\\
	\frac{\om \sqrt{2}}{\sqrt{3}}a_1 b_1 - \frac{1}{\sqrt{6}}\left(a_2 b_3+a_3 b_2\right)\\
	\frac{\om \sqrt{2}}{\sqrt{3}}a_2 b_2 - \frac{1}{\sqrt{6}}\left(a_3 b_1+a_1 b_3\right)\\
\frac{\om \sqrt{2}}{\sqrt{3}}a_3 b_3 - \frac{1}{\sqrt{6}}\left(a_1 b_2+a_2 b_1\right)
\end{matrix}\right)
\end{equation}
and
\begin{equation}\label{eq:tensorrpexp}
\boldsymbol{6}\equiv\left(\begin{matrix}\frac{1}{3}\left(a_1 b_1 + a_2 b_2 + a_3 b_3+a_2 b_3+a_3 b_2+a_3 b_1 + a_1 b_3 +a_1 b_2 + a_2 b_1\right)\\
	\frac{1}{3\sqrt{2}} a_1 b_1 - \frac{\sqrt{2}}{3} a_2 b_2 + \frac{1}{3\sqrt{2}} a_3 b_3+\frac{1}{3\sqrt{2}} (a_2 b_3+a_3 b_2) - \frac{\sqrt{2}}{3} (a_3 b_1 + a_1 b_3) + \frac{1}{3\sqrt{2}} (a_1 b_2 +a_2 b_1)\\
	\frac{1}{\sqrt{6}}\left(a_1 b_1 - a_3 b_3\right)+\frac{1}{\sqrt{6}}\left(a_2 b_3 + a_3 b_2 - a_1 b_2 - a_2 b_1\right)\\
	\frac{\sqrt{2}}{\sqrt{3}}a_1 b_1 - \frac{1}{\sqrt{6}}\left(a_2 b_3+a_3 b_2\right)\\
	\frac{\sqrt{2}}{\sqrt{3}}a_2 b_2 - \frac{1}{\sqrt{6}}\left(a_3 b_1+a_1 b_3\right)\\
\frac{\sqrt{2}}{\sqrt{3}}a_3 b_3 - \frac{1}{\sqrt{6}}\left(a_1 b_2+a_2 b_1\right)
\end{matrix}\right)
\end{equation}
respectively. On the other hand, the C-G coefficients for $\boldsymbol{\cb}$, $\boldsymbol{\ab}$ and $\boldsymbol{\bb}$ (the triplets in the RHS of Eqs.~(\ref{eq:tensor3pqr2})) are given by the same expression as the one for $\boldsymbol{\tb}$ in Eq.~(\ref{eq:tensor1exp}), i.e.~
\begin{equation}
\boldsymbol{\cb},\, \boldsymbol{\ab},\, \boldsymbol{\bb} \equiv
\left(\begin{matrix}\frac{1}{\sqrt{2}}\left(a_2 b_3 - a_3 b_2\right)\\
	\frac{1}{\sqrt{2}}\left(a_3 b_1 - a_1 b_3\right)\\
\frac{1}{\sqrt{2}}\left(a_1 b_2 - a_2 b_1\right)
\end{matrix}\right).
\end{equation}

\newpage
\begin{flalign}\label{eq:tensor18}
&\text{{\textit {\textbf {iii}}})\,\,\,}\boldsymbol{3}\otimes\boldsymbol{\bar{3}}=\boldsymbol{1}\oplus\boldsymbol{8}&
\end{flalign}
\begin{equation}\label{eq:tensor18exp}
\boldsymbol{1}\equiv \frac{1}{\sqrt{3}}\left(a_1 b_1 + a_2 b_2 + a_3 b_3\right), \quad \quad \boldsymbol{8}\equiv\left(\begin{matrix}\frac{1}{\sqrt{6}} a_1 b_1 - \sqrt{\frac{2}{3}} a_2 b_2 + \frac{1}{\sqrt{6}} a_3 b_3\\
	\frac{1}{\sqrt{2}}\left(a_1 b_1 - a_3 b_3\right)\\
	\frac{1}{\sqrt{2}}\left(a_2 b_3 + a_3 b_2\right)\\
	\frac{1}{\sqrt{2}}\left(a_3 b_1 + a_1 b_3\right)\\
	\frac{1}{\sqrt{2}}\left(a_1 b_2 + a_2 b_1\right)\\
	\frac{1}{\sqrt{2}}\left(a_2 b_3 - a_3 b_2\right)\\
	\frac{1}{\sqrt{2}}\left(a_3 b_1 - a_1 b_3\right)\\
\frac{1}{\sqrt{2}}\left(a_1 b_2 - a_2 b_1\right)
\end{matrix}\right).
\end{equation}
We define the basis for the octet representation using Eqs.~(\ref{eq:tensor18exp}). The resulting generator matrices are 
\begin{align}
&C \equiv
\left(\begin{matrix}1 & 0 & 0 & 0 & 0 & 0 & 0 & 0\\
	0 & 1 & 0 & 0 & 0 & 0 & 0 & 0\\
	0 & 0 & -\frac{1}{2} & 0 & 0 & -\frac{i \sqrt{3}}{2} & 0 & 0\\
	0 & 0 & 0 & -\frac{1}{2} & 0 & 0 & -\frac{i \sqrt{3}}{2} & 0\\
	0 & 0 & 0 & 0 & -\frac{1}{2} & 0 & 0 & -\frac{i \sqrt{3}}{2}\\
	0 & 0 & -\frac{i \sqrt{3}}{2} & 0 & 0 & -\frac{1}{2} & 0 & 0\\
	0 & 0 & 0 & -\frac{i \sqrt{3}}{2} & 0 & 0 & -\frac{1}{2} & 0\\
	0 & 0 & 0 & 0 & -\frac{i \sqrt{3}}{2} & 0 & 0 & -\frac{1}{2}
\end{matrix}\right), \quad E \equiv
\left(\begin{matrix}-\frac{1}{2} & \frac{\sqrt{3}}{2} & 0 & 0 & 0 & 0 & 0 & 0\\
	-\frac{\sqrt{3}}{2} & -\frac{1}{2} & 0 & 0 & 0 & 0 & 0 & 0\\
	0 & 0 & 0 & 1 & 0 & 0 & 0 & 0\\
	0 & 0 & 0 & 0 & 1 & 0 & 0 & 0\\
	0 & 0 & 1 & 0 & 0 & 0 & 0 & 0\\
	0 & 0 & 0 & 0 & 0 & 0 & 1 & 0\\
	0 & 0 & 0 & 0 & 0 & 0 & 0 & 1\\
	0 & 0 & 0 & 0 & 0 & 1 & 0 & 0
\end{matrix}\right),\notag \\
&V \equiv
\left(\begin{matrix}0 & 0 & \frac{1}{2\sqrt{3}} & \frac{1}{2\sqrt{3}} & \frac{1}{2\sqrt{3}} & \frac{i}{2} & \frac{i}{2} & \frac{i}{2}\\
	0 & 0 & \frac{1}{2} & \frac{1}{2} & \frac{1}{2} & -\frac{i}{2\sqrt{3}} & -\frac{i}{2\sqrt{3}} & -\frac{i}{2\sqrt{3}}\\
	\frac{1}{2\sqrt{3}} & \frac{1}{2} & \frac{2}{3} & -\frac{1}{3} & -\frac{1}{3} & 0 & 0 & 0\\
	\frac{1}{2\sqrt{3}} & \frac{1}{2} & -\frac{1}{3} & \frac{1}{6} & \frac{1}{6} & -\frac{i}{\sqrt{3}} & \frac{i}{2\sqrt{3}} & \frac{i}{2\sqrt{3}}\\
	\frac{1}{2\sqrt{3}} & \frac{1}{2} & -\frac{1}{3} & \frac{1}{6} & \frac{1}{6} & \frac{i}{\sqrt{3}} & -\frac{i}{2\sqrt{3}} & -\frac{i}{2\sqrt{3}}\\
	\frac{i}{2} & -\frac{i}{2\sqrt{3}} & 0 & -\frac{i}{\sqrt{3}} & \frac{i}{\sqrt{3}} & 0 & 0 & 0\\
	\frac{i}{2} & -\frac{i}{2\sqrt{3}} & 0 & \frac{i}{2\sqrt{3}} & -\frac{i}{2\sqrt{3}} & 0 & -\frac{1}{2} & \frac{1}{2}\\
	\frac{i}{2} & -\frac{i}{2\sqrt{3}} & 0 & \frac{i}{2\sqrt{3}} & -\frac{i}{2\sqrt{3}} & 0 & \frac{1}{2} & -\frac{1}{2}
\end{matrix}\right),\label{eq:gen8}\\
&X \equiv
\left(\begin{matrix}0 & 0 & -\frac{1}{\sqrt{3}} & \frac{1}{2\sqrt{3}} & \frac{1}{2\sqrt{3}} & 0 & -\frac{i}{2} & \frac{i}{2}\\
	0 & 0 & 0 & -\frac{1}{2} & \frac{1}{2} & \frac{i}{\sqrt{3}} & -\frac{i}{2\sqrt{3}} & -\frac{i}{2\sqrt{3}}\\
	\frac{1}{2\sqrt{3}} & -\frac{1}{2} & \frac{1}{6} & \frac{1}{6} & \frac{2}{3} & -\frac{i}{2\sqrt{3}} & \frac{i}{2\sqrt{3}} & 0\\
	-\frac{1}{\sqrt{3}} & 0 & -\frac{1}{3} & -\frac{1}{3} & \frac{1}{6} & 0 & \frac{i}{\sqrt{3}} & \frac{i}{2\sqrt{3}}\\
	\frac{1}{2\sqrt{3}} & \frac{1}{2} & -\frac{1}{3} & -\frac{1}{3} & \frac{1}{6} & -\frac{i}{\sqrt{3}} & 0 & -\frac{i}{2\sqrt{3}}\\
	-\frac{i}{2} & -\frac{i}{2\sqrt{3}} & \frac{i}{2\sqrt{3}} & -\frac{i}{2\sqrt{3}} & 0 & \frac{1}{2} & \frac{1}{2} & 0\\
	0 & \frac{i}{\sqrt{3}} & \frac{i}{\sqrt{3}} & 0 & \frac{i}{2\sqrt{3}} & 0 & 0 & -\frac{1}{2}\\
	\frac{i}{2} & -\frac{i}{2\sqrt{3}} & 0 & -\frac{i}{\sqrt{3}} & -\frac{i}{2\sqrt{3}} & 0 & 0 & -\frac{1}{2}
\end{matrix}\right).\notag\\ \notag
\end{align}
Note that for the $SU(3)$ group, the tensor product of a $\boldsymbol{3}$ and a $\boldsymbol{\tb}$ gives a $\boldsymbol{1}$ and an $\boldsymbol{8}$, i.e.~the tensor product expansion Eqs.~(\ref{eq:tensor18},~\ref{eq:tensor18exp}) is applicable to the $SU(3)$ group as well.
\begin{flalign}\label{eq:tensor288}
&\text{{\textit {\textbf {iv}}})\,\,\,}\boldsymbol{6}\otimes\boldsymbol{3}=\boldsymbol{2}\oplus\boldsymbol{8}\oplus\boldsymbol{8}&
\end{flalign}
\begin{align}\label{eq:tensor288exp}
\begin{split}
&\boldsymbol{2}\equiv\left(\begin{matrix}\frac{a_4 b_1}{\sqrt{3}}+\frac{a_5 b_2}{\sqrt{3}}+\frac{a_6 b_3}{\sqrt{3}}\\
	\frac{a_1 b_1}{3}+\frac{a_1 b_2}{3} +\frac{a_1 b_3}{3}+\frac{a_2 b_1}{3 \sqrt{2}}-\frac{\sqrt{2}a_2 b_2}{3}+\frac{a_2 b_3}{3 \sqrt{2}}+\frac{a_3 b_1}{\sqrt{6}} -\frac{a_3 b_3}{\sqrt{6}}
\end{matrix}\right),\\
&\boldsymbol{8}\equiv\left(\begin{matrix}\frac{a_1 b_1}{\sqrt{6}}-\frac{a_1 b_3}{\sqrt{6}}+\frac{a_2 b_1}{2 \sqrt{3}}-\frac{a_2 b_3}{2 \sqrt{3}}+\frac{a_3 b_1}{2}+\frac{a_3 b_3}{2}\\
	-\frac{a_1 b_1}{3 \sqrt{2}}+\frac{\sqrt{2} a_1 b_2}{3}-\frac{a_1 b_3}{3\sqrt{2}}-\frac{a_2 b_1}{6}-\frac{2 a_2 b_2}{3}-\frac{a_2 b_3}{6}-\frac{a_3 b_1}{2 \sqrt{3}}+\frac{a_3 b_3}{2 \sqrt{3}}\\
	\frac{a_1 b_2}{3 \sqrt{2}}-\frac{a_1 b_3}{3 \sqrt{2}}+\frac{a_2 b_2}{6}+\frac{a_2 b_3}{3}-\frac{a_3 b_2}{2 \sqrt{3}}-\frac{a_4 b_2}{\sqrt{3}}+\frac{a_4 b_3}{\sqrt{3}}\\
	-\frac{a_1 b_1}{3 \sqrt{2}}+\frac{a_1 b_3}{3 \sqrt{2}}-\frac{a_2 b_1}{6}+\frac{a_2 b_3}{6}+\frac{a_3 b_1}{2 \sqrt{3}}+\frac{a_3 b_3}{2 \sqrt{3}}+\frac{a_5 b_1}{\sqrt{3}}-\frac{a_5 b_3}{\sqrt{3}}\\
	\frac{a_1 b_1}{3 \sqrt{2}}-\frac{a_1 b_2}{3 \sqrt{2}}-\frac{a_2 b_1}{3}-\frac{a_2 b_2}{6}-\frac{a_3 b_2}{2\sqrt{3}}-\frac{a_6 b_1}{\sqrt{3}}+\frac{a_6 b_2}{\sqrt{3}}\\
	\frac{a_1 b_2}{3 \sqrt{2}}+\frac{a_1 b_3}{3 \sqrt{2}}+\frac{a_2 b_2}{6}-\frac{a_2 b_3}{3}-\frac{a_3 b_2}{2 \sqrt{3}}+\frac{a_4 b_2}{\sqrt{3}}+\frac{a_4 b_3}{\sqrt{3}}\\
	\frac{a_1 b_1}{3 \sqrt{2}}+\frac{a_1 b_3}{3 \sqrt{2}}+\frac{a_2 b_1}{6}+\frac{a_2 b_3}{6}-\frac{a_3 b_1}{2 \sqrt{3}}+\frac{a_3 b_3}{2 \sqrt{3}}+\frac{a_5 b_1}{\sqrt{3}}+\frac{a_5 b_3}{\sqrt{3}}\\
\frac{a_1 b_1}{3 \sqrt{2}}+\frac{a_1 b_2}{3 \sqrt{2}}-\frac{a_2 b_1}{3}+\frac{a_2 b_2}{6}+\frac{a_3 b_2}{2\sqrt{3}}+\frac{a_6 b_1}{\sqrt{3}}+\frac{a_6 b_2}{\sqrt{3}}
\end{matrix}\right), \\ &\boldsymbol{8}\equiv\left(\begin{matrix}\frac{a_4 b_1}{\sqrt{2}}-\frac{a_6 b_3}{\sqrt{2}}\\
	-\frac{a_4 b_1}{\sqrt{6}}+\frac{\sqrt{2}a_5 b_2}{\sqrt{3}} -\frac{a_6 b_3}{\sqrt{6}}\\
	-\frac{a_2 b_1}{\sqrt{2}}+\frac{a_3 b_1}{\sqrt{6}}+\frac{a_5 b_3}{\sqrt{6}}-\frac{a_6 b_2}{\sqrt{6}}\\
	-\frac{\sqrt{2} a_3 b_2}{\sqrt{3}}-\frac{a_4 b_3}{\sqrt{6}}+\frac{a_6 b_1}{\sqrt{6}}\\
	\frac{a_2 b_3}{\sqrt{2}}+\frac{a_3 b_3}{\sqrt{6}}+\frac{a_4 b_2}{\sqrt{6}}-\frac{a_5 b_1}{\sqrt{6}}\\
	\frac{2 a_1 b_1}{3}-\frac{a_2 b_1}{3 \sqrt{2}}-\frac{a_3 b_1}{\sqrt{6}}-\frac{a_5 b_3}{\sqrt{6}}-\frac{a_6 b_2}{\sqrt{6}}\\
	\frac{2 a_1 b_2}{3}+\frac{\sqrt{2} a_2 b_2}{3} -\frac{a_4 b_3}{\sqrt{6}}-\frac{a_6 b_1}{\sqrt{6}}\\
	\frac{2 a_1 b_3}{3}-\frac{a_2 b_3}{3 \sqrt{2}}+\frac{a_3 b_3}{\sqrt{6}}-\frac{a_4 b_2}{\sqrt{6}}-\frac{a_5 b_1}{\sqrt{6}}
\end{matrix}\right).
\end{split}
\end{align}
We define the basis for the doublet representation using Eqs.~(\ref{eq:tensor288exp}). The resulting generator matrices are 
\begin{equation}\label{eq:gen2}
C \equiv
\left(\begin{matrix}1 & 0\\
0 & 1
\end{matrix}\right), \quad E \equiv
\left(\begin{matrix}1 & 0\\
0 & 1
\end{matrix}\right), \quad V \equiv
i\left(\begin{matrix}-\frac{1}{\sqrt{3}} & \sqrt{\frac{2}{3}}\\
\sqrt{\frac{2}{3}} & \frac{1}{\sqrt{3}}
\end{matrix}\right), \quad X \equiv
i\left(\begin{matrix}-\frac{1}{\sqrt{3}} & \sqrt{\frac{2}{3}}\om\\
\sqrt{\frac{2}{3}}\ob & \frac{1}{\sqrt{3}}
\end{matrix}\right).
\end{equation}
In relation to the tensor product expansion Eq.~(\ref{eq:tensor288}), we may embed the irreps of $\Sigma(72\times3)$ in the irreps of $SU(3)$:
\begin{equation}
  \label{eq:orig}
    \begin{tikzpicture}[>=stealth,baseline=(current bounding box.center),anchor=base,inner sep=0pt]
      \matrix (foil) [matrix of math nodes,nodes={minimum height=1em}] {
        SU(3)&\,:\,&\boldsymbol{6}&\otimes&\boldsymbol{3}&=&\boldsymbol{10}&\oplus&\boldsymbol{8}\\
        \,\\
        \Sigma(72\times3)&\,:\,&\boldsymbol{6}&\otimes&\boldsymbol{3}&=&\boldsymbol{2}&\oplus&\boldsymbol{8}&\oplus&\boldsymbol{8}.\\
      };
\path[->] ($(foil-1-1.south)$)  edge[]     ($(foil-3-1.north)$);
\path[->] ($(foil-1-3.south)$)  edge[]     ($(foil-3-3.north)$);
\path[->] ($(foil-1-5.south)$)   edge[]    ($(foil-3-5.north)$);
\path[->] ($(foil-1-7.south)$)   edge[]    ($(foil-3-7.north)$);
\path[->] ($(foil-1-7.south)$)   edge[]    ($(foil-3-9.north)$);
\path[->] ($(foil-1-9.south)$)   edge[]    ($(foil-3-11.north)$);
    \end{tikzpicture}
  \end{equation}
\begin{flalign}\label{eq:tensor315}
&\text{{\textit {\textbf {v}}})\,\,\,}\boldsymbol{6}\otimes\boldsymbol{\tb}=\boldsymbol{3}\oplus\boldsymbol{\xb}\oplus\boldsymbol{3^p}\oplus\boldsymbol{3^q}\oplus\boldsymbol{3^r}&
\end{flalign}
\begin{equation*}
\boldsymbol{3}\equiv\left(\begin{matrix}\frac{a_1 b_1}{\sqrt{6}}+\frac{a_2 b_1}{2 \sqrt{3}}+\frac{a_3 b_1}{2}+\frac{a_5 b_3}{2}+\frac{a_6 b_2}{2}\\
	\frac{a_1 b_2}{\sqrt{6}}-\frac{a_2 b_2}{\sqrt{3}}+\frac{a_4 b_3}{2}+\frac{a_6 b_1}{2}\\
	\frac{a_1 b_3}{\sqrt{6}}+\frac{a_2 b_3}{2 \sqrt{3}}-\frac{a_3 b_3}{2}+\frac{a_4 b_2}{2}+\frac{a_5 b_1}{2}
\end{matrix}\right),
\end{equation*}
\newpage
\begin{align}
\begin{split}
&\boldsymbol{\xb}\equiv\left(\begin{matrix}-\frac{a_4 b_2}{\sqrt{6}}+\frac{a_4 b_3}{\sqrt{6}}+\frac{a_5 b_1}{\sqrt{6}}-\frac{a_5 b_3}{\sqrt{6}}-\frac{a_6 b_1}{\sqrt{6}}+\frac{a_6 b_2}{\sqrt{6}}\\
	-\frac{a_4 b_2}{2 \sqrt{3}}-\frac{a_4 b_3}{\sqrt{3}}+\frac{a_5 b_1}{2 \sqrt{3}}-\frac{a_5 b_3}{2 \sqrt{3}}+\frac{a_6 b_1}{\sqrt{3}}+\frac{a_6 b_2}{2 \sqrt{3}}\\
	\frac{a_4 b_2}{2}-\frac{a_5 b_1}{2}-\frac{a_5 b_3}{2}+\frac{a_6 b_2}{2}\\
	-\frac{a_1 b_2}{\sqrt{6}}+\frac{a_1 b_3}{\sqrt{6}}-\frac{a_2 b_2}{2 \sqrt{3}}-\frac{a_2 b_3}{\sqrt{3}}+\frac{a_3 b_2}{2}\\
	\frac{a_1 b_1}{\sqrt{6}}-\frac{a_1 b_3}{\sqrt{6}}+\frac{a_2 b_1}{2 \sqrt{3}}-\frac{a_2 b_3}{2 \sqrt{3}}-\frac{a_3 b_1}{2}-\frac{a_3 b_3}{2}\\
-\frac{a_1 b_1}{\sqrt{6}}+\frac{a_1 b_2}{\sqrt{6}}+\frac{a_2 b_1}{\sqrt{3}}+\frac{a_2 b_2}{2 \sqrt{3}}+\frac{a_3 b_2}{2}
\end{matrix}\right),\\
&\boldsymbol{3^p}\equiv\left(\begin{matrix} -\frac{a_1 b_1}{3 \sqrt{2}}-\frac{\om a_1 b_2 }{3 \sqrt{2}}-\frac{\om a_1 b_3}{3 \sqrt{2}}-\frac{a_2 b_1}{6}-\frac{\om a_2 b_2}{6}+\frac{\om a_2 b_3}{3} -\frac{a_3 b_1}{2 \sqrt{3}}+\frac{\om a_3 b_2}{2 \sqrt{3}} -\frac{\ob a_4 b_1}{\sqrt{3}} +\frac{a_5 b_3}{2 \sqrt{3}}+\frac{a_6 b_2}{2\sqrt{3}} \\
	-\frac{\om a_1 b_1}{3 \sqrt{2}}-\frac{a_1 b_2}{3 \sqrt{2}}-\frac{\om a_1 b_3}{3 \sqrt{2}}-\frac{\om a_2 b_1}{6}+\frac{a_2 b_2}{3}-\frac{\om a_2 b_3}{6} +\frac{\om a_3 b_1}{2 \sqrt{3}} -\frac{\om a_3 b_3}{2 \sqrt{3}}+\frac{a_4 b_3}{2 \sqrt{3}}-\frac{\ob a_5 b_2}{\sqrt{3}}+\frac{a_6 b_1}{2 \sqrt{3}}\\
	-\frac{\om a_1 b_1}{3 \sqrt{2}}-\frac{\om a_1 b_2}{3 \sqrt{2}}-\frac{a_1 b_3}{3 \sqrt{2}}+\frac{\om a_2 b_1}{3} -\frac{\om a_2 b_2}{6}-\frac{a_2 b_3}{6} -\frac{\om a_3 b_2}{2\sqrt{3}}+\frac{a_3 b_3}{2 \sqrt{3}}+\frac{a_4 b_2}{2 \sqrt{3}}+\frac{a_5 b_1}{2 \sqrt{3}}-\frac{\ob a_6 b_3}{\sqrt{3}} 
\end{matrix}\right),\\
&\boldsymbol{3^q}\equiv\left(\begin{matrix}-\frac{a_1 b_1}{3 \sqrt{2}}-\frac{a_1 b_2}{3 \sqrt{2}}-\frac{a_1 b_3}{3 \sqrt{2}}-\frac{a_2 b_1}{6}-\frac{a_2 b_2}{6}+\frac{a_2 b_3}{3}-\frac{a_3 b_1}{2 \sqrt{3}}+\frac{a_3 b_2}{2 \sqrt{3}}-\frac{a_4 b_1}{\sqrt{3}}+\frac{a_5 b_3}{2 \sqrt{3}}+\frac{a_6 b_2}{2 \sqrt{3}}\\
	-\frac{a_1 b_1}{3 \sqrt{2}}-\frac{a_1 b_2}{3 \sqrt{2}}-\frac{a_1 b_3}{3 \sqrt{2}}-\frac{a_2 b_1}{6}+\frac{a_2 b_2}{3}-\frac{a_2 b_3}{6}+\frac{a_3 b_1}{2 \sqrt{3}}-\frac{a_3 b_3}{2 \sqrt{3}}+\frac{a_4 b_3}{2 \sqrt{3}}-\frac{a_5 b_2}{\sqrt{3}}+\frac{a_6 b_1}{2 \sqrt{3}}\\
-\frac{a_1 b_1}{3 \sqrt{2}}-\frac{a_1 b_2}{3 \sqrt{2}}-\frac{a_1 b_3}{3 \sqrt{2}}+\frac{a_2 b_1}{3}-\frac{a_2 b_2}{6}-\frac{a_2 b_3}{6}-\frac{a_3 b_2}{2\sqrt{3}}+\frac{a_3 b_3}{2 \sqrt{3}}+\frac{a_4 b_2}{2 \sqrt{3}}+\frac{a_5 b_1}{2 \sqrt{3}}-\frac{a_6 b_3}{\sqrt{3}}
\end{matrix}\right),\\
&\boldsymbol{3^r}\equiv\left(\begin{matrix}\frac{\om a_1 b_1}{3 \sqrt{2}}+\frac{a_1 b_2}{3 \sqrt{2}}+\frac{a_1 b_3}{3\sqrt{2}}+\frac{\om a_2 b_1}{6}+\frac{a_2 b_2}{6}-\frac{a_2 b_3}{3} +\frac{\om a_3 b_1}{2 \sqrt{3}} -\frac{a_3 b_2}{2 \sqrt{3}}+\frac{\ob a_4 b_1}{\sqrt{3}}-\frac{\om a_5 b_3}{2\sqrt{3}}-\frac{\om a_6 b_2}{2 \sqrt{3}} \\
	\frac{a_1 b_1}{3 \sqrt{2}}+\frac{\om a_1 b_2}{3 \sqrt{2}}+\frac{a_1 b_3}{3 \sqrt{2}}+\frac{a_2 b_1}{6}-\frac{\om a_2 b_2}{3}+\frac{a_2 b_3}{6}  -\frac{a_3 b_1}{2 \sqrt{3}} +\frac{a_3 b_3}{2 \sqrt{3}}-\frac{\om a_4 b_3}{2 \sqrt{3}}+\frac{\ob a_5 b_2}{\sqrt{3}}-\frac{\om a_6 b_1}{2 \sqrt{3}}\\
\frac{a_1 b_1}{3 \sqrt{2}}+\frac{a_1 b_2}{3 \sqrt{2}}+\frac{\om a_1 b_3}{3 \sqrt{2}}-\frac{a_2 b_1}{3} +\frac{a_2 b_2}{6}+\frac{\om a_2 b_3}{6}+\frac{a_3 b_2}{2 \sqrt{3}}-\frac{\om a_3 b_3}{2 \sqrt{3}} -\frac{\om a_4 b_2}{2 \sqrt{3}}-\frac{\om a_5 b_1}{2 \sqrt{3}}+\frac{\ob a_6 b_3}{\sqrt{3}}
\end{matrix}\right).
\end{split}
\end{align}
The $SU(3)$ embedding corresponding to Eq.~(\ref{eq:tensor315}) is 
\begin{equation}
  \label{eq:orig}
    \begin{tikzpicture}[>=stealth,baseline=(current bounding box.center),anchor=base,inner sep=0pt]
      \matrix (foil) [matrix of math nodes,nodes={minimum height=1em}] {
        SU(3)&\,:\,&\boldsymbol{6}&\otimes&\boldsymbol{\tb}&=&\boldsymbol{3}&\oplus&\boldsymbol{15}\\
        \,\\
        \Sigma(72\times3)&\,:\,&\boldsymbol{6}&\otimes&\boldsymbol{\tb}&=&\boldsymbol{3}&\oplus&\boldsymbol{\xb}&\oplus&\boldsymbol{3^p}&\oplus&\boldsymbol{3^q}&\oplus&\boldsymbol{3^r}.\\
      };
\path[->] ($(foil-1-1.south)$)  edge[]     ($(foil-3-1.north)$);
\path[->] ($(foil-1-3.south)$)  edge[]     ($(foil-3-3.north)$);
\path[->] ($(foil-1-5.south)$)   edge[]    ($(foil-3-5.north)$);
\path[->] ($(foil-1-7.south)$)  edge[]     ($(foil-3-7.north)$);
\path[->] ($(foil-1-9.south)$)   edge[]    ($(foil-3-9.north)$);
\path[->] ($(foil-1-9.south)$)   edge[]    ($(foil-3-11.north)$);
\path[->] ($(foil-1-9.south)$)   edge[]    ($(foil-3-13.north)$);
\path[->] ($(foil-1-9.south)$)   edge[]    ($(foil-3-15.north)$);
    \end{tikzpicture}
  \end{equation}
\begin{flalign}\label{eq:tensor66}
&\text{{\textit {\textbf {vi}}})\,\,\,}\boldsymbol{6}\otimes\boldsymbol{6}=\underbrace{\boldsymbol{\xb}\oplus\boldsymbol{\xb}\oplus\boldsymbol{\xb}\oplus\boldsymbol{3}}_\text{symm}\oplus\underbrace{\boldsymbol{\xb}\oplus\boldsymbol{3^p}\oplus\boldsymbol{3^q}\oplus\boldsymbol{3^r}}_\text{antisymm}&
\end{flalign}
Here the sextet, $\boldsymbol{\xb}$, appears more than once in the symmetric part. So there is no unique way of decomposing the product space into the sum of the irreducible sextets, i.e.~the C-G coefficients are not uniquely defined. To solve this problem, we utilise the group $\Sigma(216\times3)$ which has $\Sigma(72\times3)$ as one of its subgroups. $\Sigma(216\times3)$ has three distinct types of sextets~\cite{Sigma1}, $\boldsymbol{6^x}$, $\boldsymbol{6^y}$, $\boldsymbol{6^z}$. The sextet of $\Sigma(72\times3)$ can be embedded in any of these three sextets of $\Sigma(216\times3)$. A tensor product expansion for $\Sigma(216\times3)$, equivalent to Eq.~(\ref{eq:tensor66}), is given by 
\begin{equation}\label{eq:tensor66hes}
\boldsymbol{6^x}\otimes\boldsymbol{6^x}=\underbrace{\boldsymbol{\xb^x}\oplus\boldsymbol{\xb^y}\oplus\boldsymbol{\xb^z}\oplus\boldsymbol{3^x}}_\text{symm}\oplus\underbrace{\boldsymbol{\xb^x}\oplus\boldsymbol{9}}_\text{antisymm}.
\end{equation}
In Eq.~(\ref{eq:tensor66hes}), the decomposition of the symmetric part into the irreducible sextets is unique. Hence we embed the irreps of $\Sigma(72\times3)$ in the irreps of $\Sigma(216\times3)$,
\begin{equation}\label{eq:orig}
\begin{tikzpicture}[>=stealth,baseline=(current bounding box.center),anchor=base,inner sep=0pt]
\matrix (foil) [matrix of math nodes,nodes={minimum height=1em}] {
\Sigma(216\times3)&\,:\,& \boldsymbol{6^x}&\otimes&\boldsymbol{6^x}&=&\boldsymbol{\bar{6^x}}&\oplus&\boldsymbol{\bar{6^y}}&\oplus&\boldsymbol{\bar{6^z}}&\oplus&\boldsymbol{3^x}&\oplus&\boldsymbol{\bar{6^x}}&\oplus&\boldsymbol{9}\\
\,\\
\Sigma(72\times3)&\,:\,& \boldsymbol{6}&\otimes&\boldsymbol{6}&=&\boldsymbol{\xb}&\oplus&\boldsymbol{\xb}&\oplus&\boldsymbol{\xb}&\oplus&\boldsymbol{3}&\oplus&\boldsymbol{\xb}&\oplus&\boldsymbol{3^p}&\oplus&\boldsymbol{3^q}&\oplus&\boldsymbol{3^r}.\\};
\path[->] ($(foil-1-1.south)$)  edge[]     ($(foil-3-1.north)$);
\path[->] ($(foil-1-3.south)$)  edge[]     ($(foil-3-3.north)$);
\path[->] ($(foil-1-5.south)$)   edge[]    ($(foil-3-5.north)$);
\path[->] ($(foil-1-7.south)$)  edge[]     ($(foil-3-7.north)$);
\path[->] ($(foil-1-9.south)$)   edge[]    ($(foil-3-9.north)$);
\path[->] ($(foil-1-11.south)$)  edge[]     ($(foil-3-11.north)$);
\path[->] ($(foil-1-13.south)$)   edge[]    ($(foil-3-13.north)$);
\path[->] ($(foil-1-15.south)$)  edge[]     ($(foil-3-15.north)$);
\path[->] ($(foil-1-17.south)$)   edge[]    ($(foil-3-17.north)$);
\path[->] ($(foil-1-17.south)$)  edge[]     ($(foil-3-19.north)$);
\path[->] ($(foil-1-17.south)$)   edge[]    ($(foil-3-21.north)$);
\end{tikzpicture}
\end{equation}
to obtain a unique decomposition for the case of $\Sigma(72\times3)$ as well.
\newpage

{\small
\begin{align}\label{eq:tensor66exp}
\begin{split}
&\boldsymbol{\xb}\equiv\left(\begin{matrix}\frac{2}{3}a_1 b_1-\frac{1}{3}a_2 b_2-\frac{1}{3}a_3 b_3-\frac{1}{3}a_4 b_4-\frac{1}{3}a_5 b_5-\frac{1}{3}a_6 b_6\\
	-\frac{\sqrt{2}}{3}a_2 b_2  +\frac{\sqrt{2}}{3}a_3 b_3 -\frac{1}{3 \sqrt{2}}a_4 b_4+\frac{\sqrt{2}}{3}a_5 b_5 -\frac{1}{3 \sqrt{2}}a_6 b_6-\frac{1}{3}a_{\{1} b_{2\}}\\
	-\frac{1}{\sqrt{6}}a_4 b_4+\frac{1}{\sqrt{6}}a_6 b_6 -\frac{1}{3}a_{\{1} b_{3\}}+\frac{\sqrt{2}}{3}a_{\{2} b_{3\}}\\
	-\frac{1}{3}a_{\{1} b_{4\}}-\frac{1}{3 \sqrt{2}}a_{\{2} b_{4\}}-\frac{1}{\sqrt{6}}a_{\{3} b_{4\}}+\frac{1}{\sqrt{6}}a_{\{5} b_{6\}}\\
	-\frac{1}{3}a_{\{1} b_{5\}}+\frac{\sqrt{2}}{3}a_{\{2} b_{5\}}+\frac{1}{\sqrt{6}}a_{\{4} b_{6\}}\\
	-\frac{1}{3}a_{\{1} b_{6\}}-\frac{1}{3 \sqrt{2}}a_{\{2} b_{6\}}+\frac{1}{\sqrt{6}}a_{\{3} b_{6\}}+\frac{1}{\sqrt{6}}a_{\{4} b_{5\}}
\end{matrix}\right),\\
&\boldsymbol{\xb}\equiv\left(\begin{matrix}
\frac{1}{3}(a_1 b_1+a_2 b_2+a_3 b_3)-\frac{\sqrt{2}}{3\sqrt{3}}(a_{\{1}b_{4\}}+a_{\{1}b_{5\}}+a_{\{1}b_{6\}}) +\frac{1}{6\sqrt{3}}(a_{\{2}b_{4\}}+a_{\{2}b_{6\}})-\frac{1}{3 \sqrt{3}}a_{\{2}b_{5\}}+\\
	\qquad\qquad\qquad\qquad\qquad\qquad\qquad\qquad\qquad\qquad\qquad\qquad\qquad\qquad\qquad\qquad\qquad\qquad\frac{1}{6}(a_{\{3}b_{4\}}-a_{\{3}b_{6\}})\\
	\frac{1}{3 \sqrt{2}}(-a_2 b_2+a_3 b_3)+\frac{1}{3} a_{\{1} b_{2\}}+\frac{1}{6 \sqrt{3}}(a_{\{1} b_{4\}}+a_{\{1} b_{6\}})-\frac{1}{3 \sqrt{3}}a_{\{1} b_{5\}}+\frac{\sqrt{2}}{3\sqrt{3}} (a_{\{2} b_{4\}}+a_{\{2} b_{6\}})+\\
	\qquad\qquad\qquad\qquad\qquad\qquad\qquad\qquad\qquad\qquad\qquad\qquad\qquad\qquad-\frac{1}{3 \sqrt{6}}a_{\{2} b_{5\}}+\frac{1}{3 \sqrt{2}}(-a_{\{3} b_{4\}}+a_{\{3} b_{6\}})\\
	\frac{1}{3} a_{\{1} b_{3\}}+\frac{1}{6} (a_{\{1} b_{4\}}-a_{\{1} b_{6\}})+\frac{1}{3 \sqrt{2}}(a_{\{2} b_{3\}}-a_{\{2} b_{4\}}+a_{\{2} b_{6\}})+\frac{1}{\sqrt{6}}a_{\{3} b_{5\}}\\
	\frac{\sqrt{2}}{3\sqrt{3}} \left(-a_1 b_1+a_2 b_2\right)-\frac{\sqrt{2}}{\sqrt{3}} a_4 b_4+\frac{1}{6 \sqrt{3}}a_{\{1} b_{2\}}+\frac{1}{6} a_{\{1} b_{3\}}-\frac{1}{3 \sqrt{2}} a_{\{2} b_{3\}}\\
	-\frac{\sqrt{2}}{3\sqrt{3}}a_1 b_1 -\frac{1}{3\sqrt{6}} a_2 b_2+\frac{1}{\sqrt{6}} a_3 b_3-\frac{\sqrt{2}}{\sqrt{3}} a_5 b_5-\frac{1}{3\sqrt{3}}a_{\{1} b_{2\}}\\
\frac{\sqrt{2}}{3\sqrt{3}}(-a_1 b_1+a_2 b_2)-\frac{\sqrt{2}}{\sqrt{3}} a_6 b_6+\frac{1}{6\sqrt{3}}a_{\{1} b_{2\}}-\frac{1}{6} a_{\{1} b_{3\}}+\frac{1}{3\sqrt{2}}a_{\{2} b_{3\}}
\end{matrix}\right),\\ 
&\boldsymbol{\xb}\equiv\left(\begin{matrix}\frac{1}{3 \sqrt{6}}(a_{\{1} b_{4}\}+a_{\{1} b_{5\}}+a_{\{1} b_{6\}})+\frac{1}{6 \sqrt{3}}(a_{\{2} b_{4\}}+a_{\{2} b_{6\}})-\frac{1}{3 \sqrt{3}}a_{\{2} b_{5\}}+\frac{1}{6} \left(a_{\{3} b_{4\}}-a_{\{3} b_{6\}}\right)+\\
	\qquad\qquad\qquad\qquad\qquad\qquad\qquad\qquad\qquad\qquad\qquad\qquad\qquad\qquad\qquad\qquad\qquad\frac{1}{3} \left(a_{\{4} b_{5\}}+a_{\{4} b_{6\}}+a_{\{5} b_{6\}}\right)\\
	\frac{1}{6 \sqrt{3}}(a_{\{1} b_{4\}}+a_{\{1} b_{6\}})-\frac{1}{3\sqrt{3}}a_{\{1} b_{5\}}+\frac{1}{6 \sqrt{6}}(a_{\{2} b_{4\}}+a_{\{2} b_{6\}})+\frac{\sqrt{2}}{3\sqrt{3}}a_{\{2} b_{5\}}+\frac{1}{6\sqrt{2}}(a_{\{3} b_{4\}}-a_{\{3} b_{6\}})+\\
	\qquad\qquad\qquad\qquad\qquad\qquad\qquad\qquad\qquad\qquad\qquad\qquad\qquad\qquad\qquad\qquad\frac{1}{3\sqrt{2}}(a_{\{4} b_{5\}}+a_{\{5} b_{6\}})-\frac{\sqrt{2}}{3}a_{\{4} b_{6\}}\\
	\frac{1}{6}(a_{\{1} b_{4\}}-a_{\{1} b_{6\}})+\frac{1}{6\sqrt{2}}(a_{\{2} b_{4\}}-a_{\{2} b_{6\}})+\frac{1}{2\sqrt{6}}(a_{\{3} b_{4\}}+a_{\{3} b_{6\}})+\frac{1}{\sqrt{6}}(-a_{\{4} b_{5\}}+a_{\{5} b_{6\}})\\
	-\frac{\sqrt{2}}{3\sqrt{3}}a_1 b_1-\frac{1}{3\sqrt{6}}a_2 b_2-\frac{1}{\sqrt{6}}a_3 b_3-\frac{1}{3\sqrt{3}}a_{\{1} b_{2\}}-\frac{1}{6}(a_{\{1} b_{5\}}+a_{\{1} b_{6\}})-\frac{1}{3}a_{\{1} b_{3\}}+\frac{1}{3\sqrt{2}}(-a_{\{2} b_{3\}}+a_{\{2} b_{6\}})+\\
	\qquad\qquad\qquad\qquad\qquad\qquad\qquad\qquad\qquad\qquad\qquad\qquad\qquad\qquad\qquad\qquad\qquad\quad\,\,+\frac{1}{2\sqrt{6}}a_{\{3} b_{5\}}-\frac{1}{6\sqrt{2}}a_{\{2} b_{5\}}\\
	-\frac{\sqrt{2}}{3\sqrt{3}}a_1 b_1-\frac{2\sqrt{2}}{3\sqrt{3}}a_2 b_2+\frac{2}{3\sqrt{3}}a_{\{1}b_{2\}}-\frac{1}{6}(a_{\{1}b_{4\}}+a_{\{1}b_{6\}})-\frac{1}{6\sqrt{2}}(a_{\{2}b_{4\}}+a_{\{2}b_{6\}})+\frac{1}{2\sqrt{6}}(a_{\{3}b_{4\}}-a_{\{3}b_{6\}})\\
	-\frac{\sqrt{2}}{3\sqrt{3}}a_1 b_1-\frac{1}{3\sqrt{6}}a_2 b_2-\frac{1}{\sqrt{6}}a_3 b_3-\frac{1}{3\sqrt{3}}a_{\{1} b_{2\}}+\frac{1}{3}a_{\{1} b_{3\}}-\frac{1}{6}(a_{\{1} b_{4\}}+a_{\{1} b_{5\}})+\frac{1}{3\sqrt{2}}(a_{\{2} b_{3\}}+a_{\{2} b_{4\}})+\\
	\qquad\qquad\qquad\qquad\qquad\qquad\qquad\qquad\qquad\qquad\qquad\qquad\qquad\qquad\qquad\qquad\qquad\quad-\frac{1}{6\sqrt{2}}a_{\{2} b_{5\}}-\frac{1}{2\sqrt{6}}a_{\{3} b_{5\}}
\end{matrix}\right),\\
&\boldsymbol{3}\equiv\left(\begin{matrix}\frac{1}{2\sqrt{3}}(-a_{\{1} b_{5\}}+a_{\{1} b_{6\}})-\frac{1}{2\sqrt{6}}a_{\{2} b_{5\}}-\frac{1}{\sqrt{6}}a_{\{2} b_{6\}}+\frac{1}{2\sqrt{2}}a_{\{3} b_{5\}}\\
	\frac{1}{2\sqrt{3}}(a_{\{1} b_{4\}}-a_{\{1} b_{6\}})+\frac{1}{2\sqrt{6}}(a_{\{2} b_{4\}}-a_{\{2} b_{6\}})-\frac{1}{2\sqrt{2}}(a_{\{3} b_{4\}}+a_{\{3} b_{6\}})\\
	\frac{1}{2\sqrt{3}}(-a_{\{1} b_{4\}}+a_{\{1} b_{5\}})+\frac{1}{\sqrt{6}}a_{\{2} b_{4\}}+\frac{1}{2\sqrt{6}}a_{\{2} b_{5\}}+\frac{1}{2\sqrt{2}}a_{\{3} b_{5\}}
\end{matrix}\right),\\
&\boldsymbol{\xb}\equiv\left(\begin{matrix}\frac{1}{2\sqrt{3}}(a_{[6} b_{3]}+a_{[3} b_{4]})+\frac{1}{3\sqrt{2}}(a_{[1} b_{4]}+a_{[1} b_{5]}+a_{[1} b_{6]})+\frac{1}{6}(a_{[2} b_{4]}+a_{[2} b_{6]})+\frac{1}{3}a_{[5} b_{2]}\\
	\frac{1}{2\sqrt{6}}(a_{[3} b_{4]}+a_{[6} b_{3]})+\frac{1}{6\sqrt{2}}(a_{[2} b_{4]}+a_{[2} b_{6]})+\frac{1}{3}a_{[5} b_{1]}+\frac{\sqrt{2}}{3}a_{[2} b_{5]}+\frac{1}{6}(a_{[1} b_{4]}+a_{[1} b_{6]})\\
	\frac{1}{2\sqrt{3}}(a_{[1} b_{4]}+a_{[6} b_{1]})+\frac{1}{2\sqrt{6}}(a_{[2} b_{4]}+a_{[6} b_{2]})+\frac{1}{2\sqrt{2}}(a_{[3} b_{4]}+a_{[3} b_{6]})\\
	\frac{1}{2\sqrt{3}}(a_{[1} b_{5]}+a_{[1} b_{6]})+\frac{1}{2\sqrt{6}}a_{[2} b_{5]}+\frac{1}{\sqrt{6}}a_{[6} b_{2]}+\frac{1}{2\sqrt{2}}a_{[5} b_{3]}\\
	\frac{1}{2\sqrt{3}}(a_{[1} b_{6]}+a_{[1} b_{4]})+\frac{1}{2\sqrt{6}}(a_{[2} b_{4]}+a_{[2} b_{6]})+\frac{1}{2\sqrt{2}}(a_{[3} b_{6]}+a_{[4} b_{3]})\\
	\frac{1}{2\sqrt{3}}(a_{[1} b_{4]}+a_{[1} b_{5]})+\frac{1}{2\sqrt{6}}a_{[2} b_{5]}+\frac{1}{\sqrt{6}}a_{[4} b_{2]}+\frac{1}{2\sqrt{2}}a_{[3} b_{5]}
\end{matrix}\right),\\
&\boldsymbol{3^p}\equiv\left(\begin{matrix}\frac{\om}{2\sqrt{3}}a_{[2} b_{1]}+\frac{1}{3\sqrt{2}}(a_{[6} b_{2]}+\om a_{[3} b_{2]})+\frac{1}{6}(a_{[1} b_{6]}+a_{[5} b_{1]}+\om a_{[1} b_{3]})+\frac{1}{6\sqrt{2}}a_{[5} b_{2]}+\frac{\ob}{\sqrt{6}}a_{[5} b_{6]}+\frac{1}{2\sqrt{6}}a_{[3} b_{5]}\\
	\frac{\om}{3\sqrt{2}}a_{[3} b_{2]}+\frac{1}{6}(a_{[1} b_{4]}+a_{[6} b_{1]})+\frac{\om}{3}a_{[3} b_{1]}+\frac{1}{6\sqrt{2}}(a_{[2} b_{4]}+a_{[6} b_{2]})+\frac{\ob}{\sqrt{6}}a_{[6} b_{4]}+\frac{1}{2\sqrt{6}}(a_{[4} b_{3]}+a_{[6} b_{3]})\\
	\frac{\om}{2\sqrt{3}}a_{[1} b_{2]}+\frac{1}{3\sqrt{2}}(a_{[2} b_{4]}+\om a_{[3} b_{2]})+\frac{1}{6}(a_{[1} b_{5]}+a_{[4} b_{1]}+\om a_{[1} b_{3]})+\frac{1}{6\sqrt{2}}a_{[2} b_{5]}+\frac{\ob}{\sqrt{6}}a_{[4} b_{5]}+\frac{1}{2\sqrt{6}}a_{[3} b_{5]}
\end{matrix}\right),\\
&\boldsymbol{3^q}\equiv\left(\begin{matrix}\frac{1}{2\sqrt{3}}a_{[2} b_{1]}+\frac{1}{3\sqrt{2}}(a_{[3} b_{2]}+a_{[6} b_{2]})+\frac{1}{6}(a_{[1} b_{3]}+a_{[1} b_{6]}+a_{[5} b_{1]})+\frac{1}{6\sqrt{2}}a_{[5} b_{2]}+\frac{1}{\sqrt{6}}a_{[5} b_{6]}+\frac{1}{2\sqrt{6}}a_{[3} b_{5]}\\
	\frac{1}{3\sqrt{2}}a_{[3} b_{2]}+\frac{1}{6}(a_{[1} b_{4]}+a_{[6} b_{1]})+\frac{1}{3}a_{[3} b_{1]}+\frac{1}{6\sqrt{2}}(a_{[2} b_{4]}+a_{[6} b_{2]})+\frac{1}{\sqrt{6}}a_{[6} b_{4]}+\frac{1}{2\sqrt{6}}(a_{[4} b_{3]}+a_{[6} b_{3]})\\
	\frac{1}{2\sqrt{3}}a_{[1} b_{2]}+\frac{1}{3\sqrt{2}}(a_{[2} b_{4]}+a_{[3} b_{2]})+\frac{1}{6}(a_{[1} b_{3]}+a_{[1} b_{5]}+a_{[4} b_{1]})+\frac{1}{6\sqrt{2}}a_{[2} b_{5]}+\frac{1}{\sqrt{6}}a_{[4} b_{5]}+\frac{1}{2\sqrt{6}}a_{[3} b_{5]}
\end{matrix}\right),\\
&\boldsymbol{3^r}\equiv\left(\begin{matrix}\frac{\ob}{2\sqrt{3}}a_{[2} b_{1]}+\frac{1}{3\sqrt{2}}(a_{[6} b_{2]}+\ob a_{[3} b_{2]})+\frac{1}{6}(a_{[5} b_{1]}+a_{[1} b_{6]}+\ob a_{[1} b_{3]})+\frac{1}{6\sqrt{2}}a_{[5} b_{2]}+\frac{\om}{\sqrt{6}}a_{[5} b_{6]}+\frac{1}{2\sqrt{6}}a_{[3} b_{5]}\\
	\frac{\ob}{3\sqrt{2}}a_{[3} b_{2]}+\frac{1}{6}(a_{[1} b_{4]}+a_{[6} b_{1]})+\frac{\ob}{3}a_{[3} b_{1]}+\frac{1}{6\sqrt{2}}(a_{[2} b_{4]}+a_{[6} b_{2]})+\frac{\om}{\sqrt{6}}a_{[6} b_{4]}+\frac{1}{2\sqrt{6}}(a_{[4} b_{3]}+a_{[6} b_{3]})\\
	\frac{\ob}{2\sqrt{3}}a_{[1} b_{2]}+\frac{1}{3\sqrt{2}}(a_{[2} b_{4]}+\ob a_{[3} b_{2]})+\frac{1}{6}(a_{[1} b_{5]}+a_{[4} b_{1]}+\ob a_{[1} b_{3]})+\frac{1}{6\sqrt{2}}a_{[2} b_{5]}+\frac{\om}{\sqrt{6}}a_{[4} b_{5]}+\frac{1}{2\sqrt{6}}a_{[3} b_{5]}
\end{matrix}\right).\\
\end{split}
\end{align}
}

In Eqs.~(\ref{eq:tensor66exp}) we have used the curly bracket and the square bracket to denote the symmetric sum and the antisymmetric sum respectively, i.e.~$a_{\{i} b_{j\}}=a_i b_j + a_j b_i$ and $a_{[i} b_{j]}=a_i b_j - a_j b_i$ . 

\section{Appendix B: Flavon Potentials}

Here we discuss the flavon potentials that lead to the vacuum alignments assumed in our model. The potentials we construct contain only up to the sixth-order flavon terms. It should be noted that even though our construction results in the required VEVs, we are not doing an exhaustive analysis of the most general flavon potentials involving all the possible invariant terms. However, the content we include is sufficient to realise our VEVs.

\subsection{The triplet flavons: $\phi_e$, $\phi_\mu$, $\phi_\tau$}\label{sec:triplets}

First we consider the triplet flavons $\phi_e$, $\phi_\mu$ and $\phi_\tau$. Our target is to obtain the VEVs $\langle\phi_e\rangle=\frac{i}{\sqrt{3}}(1,1,1)^T$, $\langle\phi_\mu\rangle=\frac{i}{\sqrt{3}}(1,\ob,\om)^T$ and $\langle\phi_\tau\rangle=\frac{i}{\sqrt{3}}(1,\om,\ob)^T$, Eqs.~(\ref{eq:leptvev}). The flavons $\phi_e$, $\phi_\mu$ and $\phi_\tau$ transform as $\boldsymbol{\ab}$, $\boldsymbol{\bb}$ and $\boldsymbol{\cb}$ respectively. The $3\times3$ maximal matrix $V$, Eqs.~(\ref{eq:gen3}), is one of the generators of $\boldsymbol{3}$. The corresponding generators of $\boldsymbol{\ab}$, $\boldsymbol{\bb}$ and $\boldsymbol{\cb}$ are $-V^*$, $V^*$ and $-V^*$ respectively, Eqs.~(\ref{eq:gen1p}-\ref{eq:3times1}). If the potentials of $\phi_e$, $\phi_\mu$ and $\phi_\tau$ have minima at $(-1,0,0)^T$, $(0,1,0)^T$ and $(0,0,-1)^T$, then they have minima also at $-V^*(-1,0,0)^T=\frac{i}{\sqrt{3}}(1,1,1)^T$, $V^*(0,1,0)^T=\frac{i}{\sqrt{3}}(1,\ob,\om)^T$ and $-V^*(0,0,-1)^T=\frac{i}{\sqrt{3}}(1,\om,\ob)^T$ as required. The $3\times3$ cyclic matrix $E$, Eqs.~(\ref{eq:gen3}), is another generator of $\boldsymbol{3}$ and thus of $\boldsymbol{\ab}$, $\boldsymbol{\bb}$ and $\boldsymbol{\cb}$ as well. Therefore, if the potential has minima at $(\pm1,0,0)^T$, then it has minima also at $(0,\pm1,0)^T$ and $(0,0,\pm1)^T$. In a nutshell, for obtaining the required VEVs, Eqs.~(\ref{eq:leptvev}), all we need to do is to construct potentials with minima at $(\pm1,0,0)^T$.

We start with the term $\left(\phi_e^\dagger\phi_e-1\right)^2$ which is $SU(3)$-invariant and which leads to a continuous set of minima that corresponds to unit magnitude for $\phi_e$, i.e.~$\phi_e^\dagger\phi_e=1$. Now we add terms which are invariant under $\Sigma(72\times3)$, but which break $SU(3)$ and result in a discrete set of points of minima including $(\pm1,0,0)^T$.

With two $\phi_e$ triplets which transform as $\boldsymbol{\ab}$, we may construct a conjugate sextet $\boldsymbol{\xb}$, i.e.~$\boldsymbol{\bar{3^p}}\otimes\boldsymbol{\bar{3^p}}=\boldsymbol{\bar{6}}\oplus\boldsymbol{3}$, Eqs.~(\ref{eq:tensor3pqr}). Note that the antisymmetric part, $\boldsymbol{3}$, vanishes.  With the help of Eqs.~(\ref{eq:tensor288},~\ref{eq:tensor288exp}), we combine the symmetric part, $\boldsymbol{\xb}$, with another $\phi_e$ ($\boldsymbol{\bar{3^p}}$) to obtain\footnote{Here we have omitted the distinction among $\boldsymbol{\tb}$, $\boldsymbol{\ab}$, $\boldsymbol{\ab}$ and $\boldsymbol{\ab}$. These triplets differ only with respect to a multiplication with $\pm1$. In the potential terms constructed subsequently in Sec.~(\ref{sec:triplets}), each type of triplet appears an even number of times and hence a sign flip does not have any impact.}
\begin{equation}\label{eq:tensor288conj}
\boldsymbol{\xb}\otimes\boldsymbol{\tb}=\boldsymbol{2}\oplus\boldsymbol{8}\oplus\boldsymbol{8}.
\end{equation} 
In terms of $\phi_e=(a_1,a_2,a_3)^T$, we provide the explicit expressions for the doublet,
\begin{equation}\label{eq:doubletexplicit}
\boldsymbol{2} \equiv \frac{1}{\sqrt{3}}\left(3\sqrt{2}a_1 a_2 a_3, a_1^3+a_2^3+a_3^3\right)^T,
\end{equation}
and the first octet,
\begin{equation}\label{eq:octetexplicit}
\begin{split}
\boldsymbol{8}&\equiv\frac{\sqrt{3}}{\sqrt{2}}\left(\frac{\left(a_1^3-a_3^3\right)}{\sqrt{3}}, \frac{\left(-a_1^3+2a_2^3-a_3^3\right)}{3},a_3a_2(a_3-a_2),a_1a_3(a_1-a_3),a_2a_1(a_2-a_1),\right.\\
&\qquad \qquad \qquad \qquad \qquad \qquad \qquad \qquad \left. \vphantom{\frac{1}{\sqrt{3}}}a_3a_2(a_3+a_2),a_1a_3(a_1+a_3),a_2a_1(a_2+a_1)\right)^T.\\
\end{split}
\end{equation}
It can be shown that the second octet in Eq.~(\ref{eq:tensor288conj}) is totally antisymmetric with respect to the permutation of the indices of the $\phi_e$ triplets and therefore it vanishes. Using the doublets\footnote{$\boldsymbol{2}$ as well as $\boldsymbol{8}$ are quaternionic representations. Their characters are real, but the use complex numbers can not be avoided in their representation matrices, eg.~Eqs.~(\ref{eq:gen8},~\ref{eq:gen2}). Real, complex and quaternionic representations can be identified by calculating the Frobenius-Schur indicator which gets the values $+1$, $0$ and $-1$ respectively.} from Eq.~(\ref{eq:doubletexplicit}) we obtain the invariant term $T_{e\boldsymbol{2}}\equiv\boldsymbol{2}^\dagger \boldsymbol{2}$. Similarly using the octets from Eq.~(\ref{eq:octetexplicit}) we obtain the invariant term $T_{e\boldsymbol{8}} \equiv \boldsymbol{8}^\dagger \boldsymbol{8}$.

In the previous paragraph, we obtained a conjugate sextet $\boldsymbol{\xb}$ from two $\phi_e$ triplets. With the help of Eqs.~(\ref{eq:tensor66},~\ref{eq:tensor66exp}), we take the tensor product of two of these $\boldsymbol{\xb}$s to construct three $\boldsymbol{6}$s and a $\boldsymbol{\tb}$:
\begin{equation}\label{eq:tensor66conj}
\boldsymbol{\xb}\otimes\boldsymbol{\xb}=\underbrace{\boldsymbol{6}\oplus\boldsymbol{6}\oplus\boldsymbol{6}\oplus\boldsymbol{\tb}}_{sym}\oplus\underbrace{\boldsymbol{6}\oplus\boldsymbol{\ab}\oplus\boldsymbol{\bb}\oplus\boldsymbol{\cb}}_{antisym}.
\end{equation}
Of course, the antisymmetric part vanishes. It can also be shown that the first sextet in the symmetric part is antisymmetric with respect to the permutation of the indices of the original triplets ($\phi_e$) and therefore it vanishes too. We combine the second sextet in the symmetric part with the original $\boldsymbol{\xb}$ in the LHS of Eq.~(\ref{eq:tensor66conj}) to obtain an invariant term $T_{e\boldsymbol{6}}\equiv\boldsymbol{6}^T\boldsymbol{\xb}$. $T_{e\boldsymbol{6}}$ in terms of the components of $\phi_e$ is given by\footnote{The invariant of lowest degree that breaks $SU(3)\rightarrow\Sigma(72\times3)$ as calculated in \cite{Merle} agrees with our expression for $T_{e\boldsymbol{6}}$.}:
\begin{equation}
T_{e\boldsymbol{6}}=\frac{1}{\sqrt{3}}\left(a_1^6+a_2^6+a_3^6-10a_1^3a_2^3-10a_2^3a_3^3-10a_3^3a_1^3 \right).
\end{equation} 
Note that $T_{e\boldsymbol{6}}$ is complex. A similar invariant term can be constructed using the third sextet in the RHS and the $\boldsymbol{\xb}$ in LHS. However this term, when viewed as the tensor product of the three $\boldsymbol{\xb}$s, is totally antisymmetric with respect to the permutation of indices of the $\boldsymbol{\xb}$s and thus vanishes. Note that $T_{e\boldsymbol{2}}$, $T_{e\boldsymbol{8}}$ and $T_{e\boldsymbol{8}}$ are sixth-order flavon terms. 

Combining all the non-vanishing invariant terms we write the potential
\begin{equation}\label{eq:leptpotinit}
\left(\phi_e^\dagger\phi_e-1\right)^2+k_{e1} T_{e\boldsymbol{2}} +k_{e2} T_{e\boldsymbol{8}} +k_{e3} \text{Re}\left(T_{e\boldsymbol{6}}\right)+k_{e4} \text{Im}\left(T_{e\boldsymbol{6}}\right). 
\end{equation}
By the suitable choice of coefficients $k_{e1}$, $k_{e2}$, $k_{e3}$ and $k_{e4}$, the above potential can be made to have minima at the required points, eg.~at $(\pm1,0,0)^T$. The first partial derivates at the points of extrema must be zero. To make sure that these points are minima, we do the second derivative test using the Hessian partial derivative matrix. Such a procedure has been followed in previous works eg. in~\cite{Derivative}. The first partial derivatives vanishing at $(\pm1,0,0)^T$ leads to the conditions
\begin{equation}\label{eq:leptpotfirstder}
k_{e1}+2 k_{e2} + \sqrt{3} k_{e3} = 0, \quad k_{e4}=0.
\end{equation}
Substituting Eqs.~(\ref{eq:leptpotfirstder}) in Eq.~(\ref{eq:leptpotinit}), we obtain the potential
\begin{equation}\label{eq:leptpot}
\left(\phi_e^\dagger\phi_e-1\right)^2-\left(2k_{e2}+\sqrt{3} k_{e3}\right) T_{e\boldsymbol{2}} +k_{e2} T_{e\boldsymbol{8}} +k_{e3} \text{Re}\left(T_{e\boldsymbol{6}}\right).
\end{equation}
Now we apply the second partial derivative test which gives the constraints
\begin{equation}\label{eq:leptpotcond}
k_{e2}>0, \quad k_{e3}<0.
\end{equation}
To sum up our discussion; the potential, Eq.~(\ref{eq:leptpot}), with the constraints, Eq.~(\ref{eq:leptpotcond}), has minima at $(\pm1,0,0)^T$, $(0,\pm1,0)^T$, $(0,0,\pm1)^T$ and also at $\pm\frac{i}{\sqrt{3}}(1,1,1)^T$, $\pm\frac{i}{\sqrt{3}}(1,\ob,\om)^T$, $\pm\frac{i}{\sqrt{3}}(1,\om,\ob)^T$.

Potentials similar to Eq.~(\ref{eq:leptpot}) can be written for the flavons $\phi_\mu$ and $\phi_\tau$ also, i.e.
\begin{gather}
\left(\phi_\mu^\dagger\phi_\mu-1\right)^2-\left(2k_{\mu2}+\sqrt{3} k_{\mu3}\right) T_{\mu\boldsymbol{2}} +k_{\mu2} T_{\mu\boldsymbol{8}} +k_{\mu3} \text{Re}\left(T_{\mu\boldsymbol{6}}\right)\label{eq:potleptmu}\\
\left(\phi_\tau^\dagger\phi_\tau-1\right)^2-\left(2k_{\tau2}+\sqrt{3} k_{\tau3}\right) T_{\tau\boldsymbol{2}} +k_{\tau2} T_{\tau\boldsymbol{8}} +k_{\tau3} \text{Re}\left(T_{\tau\boldsymbol{6}}\right).\label{eq:potlepttau}
\end{gather}
We need to ensure that the vacuum alignments of $\phi_e$, $\phi_\mu$ and $\phi_\tau$ are orthogonal to each other, Eqs.~(\ref{eq:leptvev}). For that purpose, we construct the cross term
\begin{equation}\label{eq:leptcross}
k_{e \mu}|\phi_e^\dagger \phi_\mu|^2 + k_{\mu \tau} |\phi_\mu^\dagger \phi_\tau|^2 + k_{\tau e} |\phi_\tau^\dagger \phi_e|^2,
\end{equation}
where $k_{e \mu}$, $k_{\mu \tau}$ and $k_{\tau e}$ are positive constants. Therefore, the complete potential for the triplet flavons is the sum of Eq.~(\ref{eq:leptpot}), Eq.~(\ref{eq:potleptmu}), Eq.~(\ref{eq:potlepttau}) and Eq.~(\ref{eq:leptcross}). Such a potential has minima at $\phi_e=(\pm1,0,0)^T$, $\phi_\mu=(0,\pm1,0)^T$ and $\phi_\tau=(0,0,\pm1)^T$, and also at $\phi_e=\pm\frac{i}{\sqrt{3}}(1,1,1)^T$, $\phi_\mu=\pm\frac{i}{\sqrt{3}}(1,\ob,\om)^T$ and $\phi_\tau=\pm\frac{i}{\sqrt{3}}(1,\om,\ob)^T$ as originally proposed.

\subsection{The sextet flavon: $\phi$}

Now we turn our attention towards constructing the potentials for the sextet flavon $\phi$. We discussed four different cases of VEVs and here the construction of potentials corresponding to all these cases are done in a rather similar framework. Let us list all the four VEVs:\footnote{Note that compared to Eqs.~(\ref{eq:vevtxmp},~\ref{eq:vevtxmm}), there is an extra negative sign in Eqs.~(\ref{eq:vevlisttxmp}, \ref{eq:vevlisttxmm}). The reason for this becomes apparent later in our discussion, but it is clear that such a choice does not have any observable consequence.}.
\begin{align}
\txm_{(\chi=+\frac{\pi}{16})}:\quad \langle\phi\rangle &= \left(\frac{-3+\sqrt{2}}{\sqrt{3}},\frac{1}{\sqrt{3}},1-\sqrt{2},0,-1,0\right),\label{eq:vevlisttxmp}\\
\txm_{(\chi=-\frac{\pi}{16})}:\quad \langle\phi\rangle &= \left(\frac{-3+\sqrt{2}}{\sqrt{3}},\frac{1}{\sqrt{3}},-1+\sqrt{2},0,-1,0\right),\label{eq:vevlisttxmm}\\
\txm_{(\phi=+\frac{\pi}{16})}:\quad \langle\phi\rangle &= \left(\frac{1+\sqrt{2}}{\sqrt{3}},\frac{1-\sqrt{2}}{\sqrt{3}}, -i \left(1-\sqrt{2}\right),0, -1+\sqrt{2}, 0\right),\label{eq:vevlisttpmp}\\
\txm_{(\phi=-\frac{\pi}{16})}:\quad \langle\phi\rangle &= \left(\frac{1+\sqrt{2}}{\sqrt{3}},\frac{1-\sqrt{2}}{\sqrt{3}}, i (1-\sqrt{2}),0, -1+\sqrt{2}, 0\right).\label{eq:vevlisttpmm}
\end{align}
These VEVs have the same magnitude, i.e.~$\langle\phi\rangle^\dagger\langle\phi\rangle=8-4\sqrt{2}$ for every given $\langle\phi\rangle$ in Eqs.~(\ref{eq:vevlisttxmp}-\ref{eq:vevlisttpmm}). Therefore we may write an $SU(3)$-invariant term $|\phi^\dagger\phi-(8-4\sqrt{2})|^2$. Just as we did in the case of the triplet flavons, here also we add terms that break $SU(3)$ but respect the $\Sigma(72\times3)$ symmetry to obtain a discrete set of minima. 

With the help of Eqs.~(\ref{eq:tensor66},~\ref{eq:tensor66exp}), We analyse the tensor product of three $\phi$s ($\boldsymbol{6}$s):
\begin{align}
\boldsymbol{6}\otimes\boldsymbol{6}\otimes\boldsymbol{6}&=\left(\underbrace{\boldsymbol{\xb}\oplus\boldsymbol{\xb}\oplus\boldsymbol{\xb}\oplus\boldsymbol{3}}_{sym}\oplus\underbrace{\xcancel{\boldsymbol{\xb}\oplus\boldsymbol{3^p}\oplus\boldsymbol{3^q}\oplus\boldsymbol{3^r}}}_{antisym}\right)\otimes\boldsymbol{6}\label{eq:tensor666a}\\
&=(\boldsymbol{\xb}\otimes\boldsymbol{6})\oplus(\boldsymbol{\xb}\otimes\boldsymbol{6})\oplus(\boldsymbol{\xb}\otimes\boldsymbol{6})\oplus(\boldsymbol{3}\otimes\boldsymbol{6})\label{eq:tensor666b}.
\end{align}
Each tensor product $(\boldsymbol{\xb}\otimes\boldsymbol{6})$ in Eq.~(\ref{eq:tensor666b}) contains the invariant term $\boldsymbol{\xb}^T\boldsymbol{6}$. Such invariants constructed from the first $(\boldsymbol{\xb}\otimes\boldsymbol{6})$ and the second $(\boldsymbol{\xb}\otimes\boldsymbol{6})$ in Eq.~(\ref{eq:tensor666b}) are named $T_{\boldsymbol{6}}$ and $T'_{\boldsymbol{6}}$ respectively. In terms of the components of the sextet flavon, $\phi=(a_1,a_2,a_3,a_4,a_5,a_6)^T$, they are given by

\begin{align}\label{eq:t6invar}
\begin{split}
T_{\boldsymbol{6}} &=\frac{2}{3}a_1^3-\frac{\sqrt{2}}{3}a_2^3-a_1\left(a_2^2+a_3^2+a_4^2+a_5^2+a_6^2\right)\\
&\qquad\qquad+\sqrt{2}a_2\left(a_3^2+a_5^2\right)-\frac{1}{\sqrt{2}}a_2\left(a_4^2+a_6^2\right)-\frac{\sqrt{3}}{\sqrt{2}}a_3\left(a_4^2+a_6^2\right)+\sqrt{6} a_4 a_5 a_6,
\end{split}\\
\begin{split}
T'_{\boldsymbol{6}}&=\frac{1}{3}a_1^3 -\frac{1}{3\sqrt{2}}a_2^3-\frac{\sqrt{2}}{\sqrt{3}}\left(a_4^3+a_5^3+a_6^3\right)+a_1\left(a_2^2+a_3^2\right)+\frac{1}{\sqrt{2}}a_2 a_3^2\\
&\qquad\qquad-\frac{\sqrt{2}}{\sqrt{3}}a_1^2\left(a_4-a_5+a_6\right)+\frac{\sqrt{2}}{\sqrt{3}}a_2^2\left(a_4+a_6\right)-\frac{1}{\sqrt{6}}a_2^2a_5+\frac{\sqrt{3}}{\sqrt{2}}a_3^2 a_5\\
&\qquad\qquad+a_1 a_3 \left(a_4-a_6\right) + \sqrt{2}a_2 a_3\left(a_6-a_4\right) + \frac{1}{\sqrt{3}}a_1 a_2\left(a_6+a_4\right) +\frac{2}{\sqrt{3}} a_1 a_2 a_5.
\end{split}
\end{align}

The invariant term constructed from the third $(\boldsymbol{\xb}\otimes\boldsymbol{6})$ in Eq.~(\ref{eq:tensor666b}) is totally antisymmetric under the permutation of the indices of the three $\boldsymbol{6}$s in the LHS of Eq.~(\ref{eq:tensor666a}) and therefore it vanishes. 

Consider the tensor product space $\phi\otimes\phi\otimes\phi$. Any specific alignment of $\phi$, eg.~the VEV $\langle\phi\rangle$, has a corresponding alignment in the tensor product space, eg.~$\langle\phi\rangle\otimes\langle\phi\rangle\otimes\langle\phi\rangle$. We have shown that the tensor product space $\phi\otimes\phi\otimes\phi$ contains two non-vanishing invariant directions, the ones that correspond to $T_{\boldsymbol{6}}$ and $T'_{\boldsymbol{6}}$. We take the projection of the alignment $\langle\phi\rangle\otimes\langle\phi\rangle\otimes\langle\phi\rangle$ along these directions and call them $\langle\phi\rangle\otimes\langle\phi\rangle\otimes\langle\phi\rangle_{T_{\boldsymbol{6}}}$ and $\langle\phi\rangle\otimes\langle\phi\rangle\otimes\langle\phi\rangle_{T'_{\boldsymbol{6}}}$ respectively. It can be shown that, for every VEV given in Eqs.~(\ref{eq:vevlisttxmp}-\ref{eq:vevlisttpmm})\footnote{The reason for changing the signs of Eqs.~(\ref{eq:vevlisttxmp},~\ref{eq:vevlisttxmm}) compared to Eqs.~(\ref{eq:vevtxmp},~\ref{eq:vevtxmm}) was to ensure that for all the VEVs, Eqs.~(\ref{eq:vevlisttxmp}-\ref{eq:vevlisttpmm}), the values of $\langle\phi\rangle\otimes\langle\phi\rangle\otimes\langle\phi\rangle_{T_{\boldsymbol{6}}}$ (and $\langle\phi\otimes\langle\phi\rangle\otimes\langle\phi\rangle_{T'_{\boldsymbol{6}}}$) have the same sign.}, we get $\langle\phi\rangle\otimes\langle\phi\rangle\otimes\langle\phi\rangle_{T_{\boldsymbol{6}}}=\sqrt{3}$ and $\langle\phi\rangle\otimes\langle\phi\rangle\otimes\langle\phi\rangle_{T'_{\boldsymbol{6}}}=\sqrt{3}\left(5\sqrt{2}-7\right)$. In other words $|T_{\boldsymbol{6}}-\sqrt{3}|^2$ and $|T'_{\boldsymbol{6}}-\sqrt{3}\left(5\sqrt{2}-7\right)|^2$ have the minimum value zero when the flavon field acquires any of the four VEVs. Thus we construct the flavon potential
\begin{equation}\label{eq:commonpot}
\left|\phi^\dagger\phi-(8-4\sqrt{2})\right|^2 + k_{\nu1} \left|T_{\boldsymbol{6}}-\sqrt{3}\right|^2 + k_{\nu2} \left|T'_{\boldsymbol{6}}-\sqrt{3}\left(5\sqrt{2}-7\right)\right|^2
\end{equation}
where $k_{\nu 1}$ and $k_{\nu 2}$ are positive constants. This potential, Eq.(\ref{eq:commonpot}), has minima at the required VEVs, Eqs.~(\ref{eq:vevlisttxmp}-\ref{eq:vevlisttpmm}), as required. Further analysis shows that the points of minima are not discrete, but rather they form a continuous set. In order to remove this ambiguity and ensure a discrete set of minima we add more invariant terms to the potential. We construct such terms by coupling the triplet flavons $\phi_e$, $\phi_\mu$ and $\phi_\tau$ with sextet flavon $\phi$.

First we construct sextets ($\boldsymbol{\xb}$s) by combining two triplets ($\boldsymbol{\tb}$s) using the conjugate forms of Eqs.~(\ref{eq:tensor3pqr},~\ref{eq:tensor3pqr2}). The various possiblilities are $\phi_e\otimes \phi_e$, $\phi_\mu\otimes \phi_\mu$, $\phi_\tau\otimes \phi_\tau$, $\phi_\mu\otimes \phi_\tau$, $\phi_\tau\otimes \phi_e$ and $\phi_e\otimes \phi_\mu$. The sextets so constructed are combined with the sextet flavon $\phi$ to obtain invariants, namely $T_{ee}$, $T_{\mu\mu}$, $T_{\tau\tau}$, $T_{\mu\tau}$, $T_{\tau e}$ and $T_{e\mu}$. In the tensor product space $\phi_\alpha\otimes \phi_\beta \otimes \phi$ where $\alpha, \beta = e, \mu, \tau$, we consider the specific alignment $\langle\phi_\alpha\rangle\otimes \langle\phi_\beta\rangle \otimes \langle\phi\rangle$ which corresponds to the required VEVs of the flavons given in Eqs.~(\ref{eq:leptvev},~\ref{eq:vevlisttxmp}-\ref{eq:vevlisttpmm}). As was done previously, we take the projection of this alignment along the direction of the invariant $T_{\alpha\beta}$, i.e.~$\langle\phi_\alpha\rangle \otimes \langle\phi_\beta\rangle \otimes \langle\phi\rangle_{T_{\alpha\beta}}$. Finally we construct the potential term
\begin{equation}\label{eq:uniquepot} 
\displaystyle \sum_{\alpha,\beta} k'_{\alpha\beta} \left|T_{\alpha\beta}-\langle\phi_\alpha\rangle \otimes \langle\phi_\beta\rangle \otimes \langle\phi\rangle_{T_{\alpha\beta\nu}}\right|^2
\end{equation}
where $k'_{\alpha\beta}$ are positive constants and the summation is over $(\alpha,\beta)=$ $(e,e)$, $(\mu,\mu)$, $(\tau,\tau)$, $(\mu,\tau)$, $(\tau,e)$ and $(e,\mu)$. The values of $\langle\phi_\alpha\rangle \otimes \langle\phi_\beta\rangle \otimes \langle\phi\rangle_{T_{\alpha\beta}}$ corresponding to the VEVs, Eqs.~(\ref{eq:vevlisttxmp}-\ref{eq:vevlisttpmm}), are given in Table~\ref{tab:vev}. 

\begin{table}[]
\begin{center}
\begin{tabular}{||c||c|c|c|c||}
    \hline \hline
    \multirow{2}{*}{} & \multicolumn{4}{c||}{$\langle\phi_\alpha\rangle \otimes \langle\phi_\beta\rangle \otimes \langle\phi\rangle_{T_{\alpha\beta\nu}}$} \\
    				    \cline{2-5}
    $(\alpha,\beta)$&$\txm_{(\chi=+\frac{\pi}{16})}$& $\txm_{(\chi=-\frac{\pi}{16})}$& $\tpm_{(\phi=+\frac{\pi}{16})}$& $\tpm_{(\phi=-\frac{\pi}{16})}$\\
    &({\small Eq.~(\ref{eq:vevlisttxmp})}) & ({\small Eq.~(\ref{eq:vevlisttxmm})}) & ({\small Eq.~(\ref{eq:vevlisttpmp})}) & ({\small Eq.~(\ref{eq:vevlisttpmm})}) \\
    \hline \hline
    $(e,e)$		&$1$ 	       & $1$         & $-1$         & $-1$           \\
    \hline 
    $(\mu,\mu)$		&$\frac{1-\sqrt{2}}{2}+i\frac{1+\sqrt{2}}{2\sqrt{3}}$  	& $-\frac{1}{2}+i\frac{-1+2\sqrt{2}}{2\sqrt{3}}$          & $\frac{(-1+\sqrt{2})(1+\sqrt{3})(1-i\sqrt{3})}{2\sqrt{6}}$          & $\frac{(-1+\sqrt{2})(-1+\sqrt{3})(1-i\sqrt{3})}{2\sqrt{6}}$           \\
    \hline 
    $(\tau,\tau)$	& $\frac{1-\sqrt{2}}{2}-i\frac{1+\sqrt{2}}{2\sqrt{3}}$	&   $-\frac{1}{2}-i\frac{-1+2\sqrt{2}}{2\sqrt{3}}$         &  $\frac{(-1+\sqrt{2})(-1+\sqrt{3})(1+i\sqrt{3})}{2\sqrt{6}}$    & $\frac{(-1+\sqrt{2})(1+\sqrt{3})(1+i\sqrt{3})}{2\sqrt{6}}$           \\
    \hline 
    $(\mu,\tau)$	&$\frac{3-\sqrt{2}}{2\sqrt{3}}-i\frac{1}{2}$ 	       	& $\frac{3-\sqrt{2}}{2\sqrt{3}}-i\frac{1}{2}$          & $-\frac{1+\sqrt{2}}{2\sqrt{3}}+i\frac{1}{2}$          & $-\frac{1+\sqrt{2}}{2\sqrt{3}}+i\frac{1}{2}$           \\
    \hline
    $(\tau,e)$		&$-\frac{1}{2\sqrt{6}}+i\frac{1}{2\sqrt{2}}$ 	       	& $-\frac{1}{2\sqrt{6}}+i\frac{1}{2\sqrt{2}}$          & $\frac{-1+\sqrt{2}}{2\sqrt{6}}-i\frac{-1+\sqrt{2}}{2\sqrt{2}}$          & $\frac{-1+\sqrt{2}}{2\sqrt{6}}-i\frac{-1+\sqrt{2}}{2\sqrt{2}}$           \\
    \hline 
    $(e,\mu)$		&$\frac{-1+2\sqrt{2}}{2\sqrt{6}}-i\frac{1}{2\sqrt{2}}$ 	& $\frac{3-2\sqrt{2}}{2\sqrt{6}}-i\frac{1}{2\sqrt{2}}$          & $\frac{1-\sqrt{2}}{2\sqrt{6}}+i\frac{(-1+\sqrt{2})(3-2\sqrt{3})}{6\sqrt{2}}$   & $\frac{1-\sqrt{2}}{2\sqrt{6}}+i\frac{(-1+\sqrt{2})(3+2\sqrt{3})}{6\sqrt{2}}$           \\
    \hline \hline
  \end{tabular}
\end{center}
\caption{The projection of the tensor products of VEVs along the corresponding invariant directions.}
\label{tab:vev}
\end{table}

The terms given in Eq.~(\ref{eq:commonpot}) and Eq.~(\ref{eq:uniquepot}) form a potential with a discrete set of minima which includes the requied VEV, Eqs.~(\ref{eq:vevlisttxmp}-\ref{eq:vevlisttpmm}). Note that, we use highly specific constants in our potential, eg. $(8-4\sqrt{2})$, $\sqrt{3}$ and $\sqrt{3}\left(5\sqrt{2}-7\right)$ in Eq.~(\ref{eq:commonpot}). Put another way, we may be able to construct any given mass matrix by suitably tweaking such constants. This situation can be made a lot less arbitrary by imposing additional symmetries on top of $\Sigma(72\times3)$. In this context, we can not help noticing the appearance of $\frac{\pi}{16}$ and the factor $-1+\sqrt{2}$ ($=\tan \frac{\pi}{8}$) throughout this paper. They give hints towards the presence of additional symmetries like $Z_{16}$. Such topics are beyond the scope of this paper, but will be discussed in a future publication.

\providecommand{\href}[2]{#2}\begingroup\raggedright\endgroup

\end{document}